\documentclass[12pt]{article}
\usepackage{graphicx,amssymb,amsmath}

\newcommand{\be}{\begin{equation}}
\newcommand{\ee}{\end{equation}}
\newcommand{\bea}{\begin{eqnarray}}
\newcommand{\eea}{\end{eqnarray}}
\newcommand{\beas}{\begin{eqnarray*}}
\newcommand{\eeas}{\end{eqnarray*}}

\newcommand{\x}{{\bf x}}

\newcommand{\hi}{{\hat{\imath}}}
\newcommand{\hj}{{\hat{\jmath}}}

\begin{document}
\begin{titlepage}

\begin{center}

{\Large Decoding the hologram: Scalar fields interacting with gravity}

\vspace{8mm}

\renewcommand\thefootnote{\mbox{$\fnsymbol{footnote}$}}
Daniel Kabat${}^{1}$\footnote{daniel.kabat@lehman.cuny.edu},
Gilad Lifschytz${}^{2}$\footnote{giladl@research.haifa.ac.il}

\vspace{4mm}

${}^1${\small \sl Department of Physics and Astronomy} \\
{\small \sl Lehman College, City University of New York, Bronx NY 10468, USA}

\vspace{2mm}

${}^2${\small \sl Department of Mathematics and Physics} \\
{\small \sl University of Haifa at Oranim, Kiryat Tivon 36006, Israel}

\end{center}

\vspace{8mm}

\noindent
We construct smeared CFT operators which represent a scalar field in
AdS interacting with gravity.  The guiding principle is
micro-causality: scalar fields should commute with themselves at
spacelike separation.  To ${\cal O}(1/N)$ we show that a correct and
convenient criterion for constructing the appropriate CFT operators is
to demand micro-causality in a three-point function with a boundary
Weyl tensor and another boundary scalar.  The resulting bulk
observables transform in the correct way under AdS isometries and
commute with boundary scalar operators at spacelike separation, even
in the presence of metric perturbations.
\end{titlepage}
\setcounter{footnote}{0}
\renewcommand\thefootnote{\mbox{\arabic{footnote}}}

\section{Introduction}
A long-standing problem in formulating any theory of quantum gravity
is to specify the observables of the theory.  For a review of various
approaches see \cite{Isham:1992ms}.  The problem becomes particularly
acute in the context of the AdS/CFT correspondence
\cite{Maldacena:1997re}, where correlation functions of local
operators in the boundary CFT in principle provide a complete set of
observables.  The challenge is then deciding how local or semi-local
observables in the bulk can be expressed in terms of the CFT.

The purpose of the present paper is to construct a set of smeared or
non-local observables in the CFT which can be used to represent a
scalar field in the bulk interacting with gravity.  The approach we
take builds on a long series of developments.  To summarize the
history, CFT operators which represent free fields in the bulk were constructed in \cite{Banks:1998dd,Balasubramanian:1998sn,Dobrev:1998md,Bena:1999jv,Hamilton:2005ju,Hamilton:2006az,Hamilton:2006fh} for scalar fields and in \cite{Heemskerk:2012mq,Kabat:2012hp} for fields with spin.  These constructions proceeded essentially by solving free wave equations in AdS.  The resulting observables can be
used to describe bulk physics in the large $N$ limit of the CFT.  The
corrections needed to account for bulk interactions, corresponding to
the $1/N$ expansion of the CFT, were constructed in \cite{Kabat:2011rz} and further developed in
\cite{Heemskerk:2012mn}.  In particular
\cite{Kabat:2011rz} advocated an approach based on bulk
micro-causality and argued that the correct bulk observables could be
built up order-by-order in the $1/N$ expansion, by demanding that they
obey appropriate commutation relations at spacelike separation.  This
is the approach we will take in the present paper.  Thus our work
builds on the free-field representations of \cite{Kabat:2012hp} and
takes the approach to including interactions developed in
\cite{Kabat:2011rz}.  Our analysis parallels the study of charged
scalar fields interacting with gauge fields carried out in
\cite{Kabat:2012av}.

Since this work culminates a long series of developments, the rest of
the introduction is devoted to a survey of the general approach,
including the issues it addresses and the insights it has to offer.
Readers familiar with the approach who wish to get to the technical
details may skip to section 2.

\bigskip

In any approach to quantum gravity one would like to understand how
local or semi-local observables arise from the formalism of the
underlying theory.  One can of course choose a covariant gauge where
all fields look local \cite{DeWitt:1967ub}.  But in a covariant gauge
the operators, although local, are not necessarily physical.  On the
other hand in the canonical quantization of general relativity coupled
to matter via the ADM formalism
\cite{Arnowitt:1962hi,Arnowitt:1960zza} one can show that in many nice
gauges matter field operators obey the usual equal-time commutators
with themselves and with some (but not all) of the gravity degrees of
freedom.  It's easy to see that matter field operators cannot commute
at spacelike separation with all of the gravity degrees of
freedom. For example if there is a boundary at spatial infinity, then
the Hamiltonian is a boundary term, but its commutator with matter
fields should still generate a time translation.  This means that
matter fields are not in fact local operators: rather they're
non-local operators in the bulk.\footnote{At best one could call them
quasi-local.  The matter fields we will construct can be thought of
as local operators in the bulk attached to Wilson lines that run off
to infinity.}  This can also be seen from the fact that physical
operators must commute with the constraints. The constraints in
quantum gravity generate diffeomorphisms. No locally defined quantity
can commute with the constraints, since there is no local way of
defining the position of the operator. However if there is a boundary
at infinity then the operators and coordinates on the boundary are
diffeomorphism invariant, since as usual the constraints are
implemented with fall-off conditions at infinity. One can use these
boundary coordinates to define an invariant position in the bulk, by
following geodesics in from the boundary for some given proper length.

If there is a boundary, then the Hamiltonian of the bulk theory is
non-zero and one can talk about unitarity. At the boundary there are
always local observables, and since the Hamiltonian is a boundary
operator, one can ask if the theory is unitary for the boundary
operators by themselves or if one needs the bulk operators as well.
The AdS/CFT conjecture states that the boundary operators by
themselves form a unitary theory.\footnote{For an argument that this
must be the case in quantum gravity see \cite{Marolf:2008mf}.}  But
this can only work if the boundary operators are the only operators in
the theory.  This means that the bulk operators, which are usually
thought of as describing the actual spacetime physics, can be
expressed in terms of the boundary operators. This is a drastic
(holographic) reduction in the number of degrees of freedom, but it
seems to be forced on us by unitarity.

The program of building bulk operators amounts to constructing the map
between the non-local bulk observables and the boundary
observables. It is important to stress that all information and
degrees of freedom live at the boundary, and that the notion of the
bulk spacetime is emergent. There is no need to introduce bulk
operators for the theory to be well-defined or complete. In fact there
is no ``good'' or ``bad'' definition of bulk operators, until we have
specified what properties we wish these operators to have. The
properties one wants the bulk operators to have are tied to what we
think the notion of a spacetime means, and this we can only define
through our experience with semiclassical physics. However one cannot
expect that a semiclassical spacetime will emerge for any state of the
boundary theory. Indeed a necessary requirement is that it obeys $1/N$
factorization. As such the most reasonable starting point to define
bulk operators is to look for a map in situations where we know a
semiclassical spacetime emerges,\footnote{For some speculation on the
definition of bulk operators in a more general framework see
\cite{Lifschytz:2000bj} and section 6 of \cite{Kabat:2012hp}.} such
as a CFT with large central charge in its vacuum state, which is dual
to an empty AdS space. A minimal requirement on the bulk operators is
that they will obey micro-causality with respect to the bulk causal
structure. One might also hope that they will describe bulk fields
which transform correctly under AdS isometries, and that they will
have the appropriate local interactions expected for fields in the
bulk. It turns out the above three requirements are tied together in
an interesting way.

Assuming the boundary theory is a CFT, with a central charge that we
will label by $N$, then in the large $N$ limit CFT correlators
factorize into a product of two-point functions.  In this case the
boundary-to-bulk map was constructed in the early days of AdS/CFT. The
map can be written as a sum over modes in momentum space
\cite{Banks:1998dd,Balasubramanian:1998sn}, or as a smearing function
over the entire boundary \cite{Bena:1999jv}, or just the part of the
boundary which is spacelike to the bulk point
\cite{Hamilton:2005ju,Hamilton:2006az}, or over a compact region on
the complexified boundary \cite{Hamilton:2006fh}. The key ingredient
is that in the large central charge limit, commutators of operators
are c-numbers, and after normalization the operator Fourier components
can serve as creation and annihilation operators that describe free
fields in the bulk.  However conventional bulk perturbation theory,
which is based on creation and annihilation operators, cannot simply
be adopted to take interactions into account.  The problem is that in
$1/N$ perturbation theory the operators will no longer commute to a
c-number, so the identification of creation and annihilation operators
with operator Fourier components breaks down. Indeed if one tries to
use the zeroth-order smearing function to compute a bulk three-point
function the result, while AdS covariant, does not respect bulk
micro-causality. It turns out there are two equivalent approaches to
constructing bulk operators that respect micro-causality
\cite{Kabat:2011rz}. One approach, which is simple to implement if one
knows the bulk action, is to solve the bulk equations of motion in
perturbation theory. This is most conveniently done with the help of a
spacelike Greens function
\cite{Hamilton:2005ju,Hamilton:2006az,Heemskerk:2012mn}. Another
approach, more intrinsic to the CFT, is to correct the zeroth-order
definition of the bulk operator by adding appropriately-smeared
higher-dimension operators.  One fixes the coefficients in front of
these higher-dimension operators by requiring bulk micro-causality in
all three-point functions.  This procedure was carried out for scalars
in \cite{Kabat:2011rz,Kabat:2012av}, where it was found that for bulk
scalar fields one needs to add an infinite tower of smeared
higher-dimension primary scalar operators. This construction is
possible in $1/N$ perturbation theory, where the needed primary
scalars are constructed as multi-trace operators in the CFT.  The two
approaches to constructing bulk operators are equivalent.  In both
approaches what one is constructing is the bulk Heisenberg picture
field operator.

It is important to realize that the required tower of higher-dimension
primary scalars is only guaranteed to exist in $1/N$ perturbation
theory.  Most likely it does not exist in a unitary CFT with finite
central charge. Thus even at this level having finite central charge
precludes micro-causality of bulk operators. Note that the breakdown
of locality associated with a finite Planck length in the bulk does
not manifest itself in correlators as the absence of a singularity at
lightlike-separated or coincident points, but rather through a
breakdown of micro-causality. This may seem strange, since we have not
yet incorporated any gravitational degrees of freedom.  But note that
this breakdown comes from demanding that the boundary operators are
described by a consistent unitary CFT.  This is certainly not the case
for scalar fields propagating on a fixed background, where the
boundary data does not evolve in a unitary way by itself.  Another
point we wish to stress is that in this construction, the operators
needed to correct the zeroth order definition of a bulk field are
smeared primary scalars in the CFT. It may seem obvious that one
should smear primary scalars, but as we will see, this is in fact a
consequence of demanding micro-causality.  Moreover, smeared primary
scalars transform like a local scalar field in the bulk under AdS
isometries.  So we see that both the transformation properties in the
bulk (the fact that the bulk operator transforms like a scalar field),
and the emergence of local bulk interactions (via the addition of
multi-trace operators in the CFT), are a consequence of demanding bulk
micro-causality.

The construction for interacting scalar fields was extended to the
case of charged scalar fields interacting with a bulk gauge field in
\cite{Kabat:2012av}.  Since the bulk theory has a gauge redundancy,
the first step is fixing a gauge.  A natural choice is holographic
gauge, which sets the radial component of the gauge field to zero.
Solving the bulk equations of motion in holographic gauge is
relatively straightforward.  However the micro-causality conditions in
holographic gauge are somewhat complicated. Canonical quantization of
this system shows that, although matter fields obey canonical
commutators with themselves, they have non-zero (and non-local)
commutators with certain components of the gauge field.  These matter
-- gauge commutators do not vanish, even at spacelike
separation.\footnote{Similar non-local commutators arise in
electrodynamics in Coulomb gauge.  They are required by the Gauss
constraint, which allows the total charge to be expressed as a
surface integral at infinity.}  These non-trivial commutators make
the CFT approach to constructing bulk operators more subtle, since one
cannot simply demand that operators commute at spacelike separation in
the bulk.

From the CFT perspective, the obstruction to building commuting
observables can be traced back to the Ward identities associated with
the conserved current.  Consider, for example, the CFT three-point
function $\langle {\cal O} \bar{\cal O} j_\mu \rangle$ of a charged
scalar primary ${\cal O}$ and its complex conjugate $\bar{\cal O}$
with a conserved current $j_\mu$.  If one smears ${\cal O}$ into the
bulk using the zeroth-order smearing function, the resulting mixed
bulk -- boundary correlator $\langle \phi \bar{\cal O} j_\mu \rangle$
violates micro-causality.  A straightforward attempt to restore
micro-causality, by adding smeared higher-dimension primary operators
to the definition of the bulk field $\phi$, goes nowhere: due to the
Ward identities of the CFT, the addition of such smeared
higher-dimension primaries cannot change the correlator.

Fortunately one can proceed indirectly, and demand micro-causality in
a three-point function $\langle \phi \bar{\cal O} F_{\mu\nu}\rangle$
involving the boundary field strength $F = d j$.  Micro-causality can
be restored by adding smeared operators to the definition of $\phi$
which are higher-dimension and multi-trace but are not
primary.\footnote{It is the absence of the divergence of the current
as an operator, i.e.\ $\partial_\mu j^\mu = 0$, which prevents one
from building a primary scalar.  With non-conserved currents there
is no obstruction to constructing a primary scalar
\cite{Kabat:2012hp}.}  Since the operators we add are not smeared
primaries, we obtain a bulk operator which does not transform like a
scalar under AdS isometries.  Rather, in the example worked out in
\cite{Kabat:2012av}, the bulk operator transforms exactly like a
charged scalar field in the bulk attached to a Wilson line that runs
off to infinity. This shows that by demanding an appropriate statement
of micro-causality in the bulk, one automatically obtains operators
with the correct transformation properties under bulk isometries.  We
find this connection rather nice.

The purpose of the present paper is to generalize this construction to
the case of scalar fields interacting with gravity. In section
\ref{sect:bulk} we start with a canonical treatment of a scalar field
coupled to gravity in the ADM formalism. We do this because the
commutation relations are gauge-dependent, and although ADM showed
(among many other things) that matter Poisson brackets often have the
standard form, holographic gauge was not among the class of gauges
they considered.  So we need to check if matter fields commute at
spacelike separation in holographic gauge, which turns out to be true
due to conservation of the matter stress tensor.  In section
\ref{sect:gauge} we turn to the CFT construction and revisit the
$U(1)$ vector case, extending some of the results of
\cite{Kabat:2012av}. In particular we show that matter operators
commute with each other at spacelike separation, even in the presence
of a gauge field. In section \ref{sect:gravity} we consider scalar
fields interacting with gravity. The CFT construction is carried out
for dimensions $d \geq 4$, where the analysis is facilitated by the
existence of a boundary Weyl tensor (constructed from the stress
tensor of the CFT) with four distinct indices.  We show how to
construct a bulk scalar field which obeys micro-causality inside a
three-point correlation function with a boundary scalar and a boundary
Weyl tensor.  As a result the bulk scalar commutes at spacelike
separation with other matter fields, but not with all components of
the metric.  This is expected, given that the Hamiltonian can be
written as a surface integral.  Again the construction involves adding
smeared operators which are higher-dimension and multi-trace but are
not primary.  Somewhat remarkably, just as in the gauge field case,
the transformation of such operators under AdS isometries exactly
matches the transformation expected for a bulk scalar field in
holographic gauge.  Some background computations are given in the
appendices.

\section{Bulk perspective\label{sect:bulk}}

Our main interest in this paper will be to take a boundary perspective.  We want to construct operators in the CFT which can mimic local observables in the bulk.  Our guiding principle in the construction
will be to enforce an appropriate notion of bulk micro-causality.  That is, we propose that the algebra of local operators in the CFT can be lifted to an algebra of local operators in the bulk, roughly speaking
by requiring that bulk operators commute at spacelike separation.  In pursuing this program there are two issues which confront us.
\begin{itemize}
\item
The bulk fields have gauge redundancy, associated with diffeomorphisms and other gauge symmetries that act trivially on the CFT.  To construct bulk observables these gauge symmetries must somehow
be fixed.  What choice of gauge should we make?
\item
In a gauge theory the commutators are gauge-dependent and can be non-zero at spacelike separation.  So we need a refined statement of bulk micro-causality, appropriate to our choice of gauge.  What are the correct commutation relations to impose on our bulk observables?
\end{itemize}
It's useful to gain some insight into these issues before jumping into the CFT construction.  So in this section we take a bulk perspective, and study commutators and gauge fixing for a theory of gravity coupled to matter in AdS.

Of course gravity plus matter has been studied before, in particular in \cite{Arnowitt:1960zza} and section 6.2 of \cite{Arnowitt:1962hi}.  In these works it
was indeed found that matter fields continue to obey canonical brackets even when coupled to gravity.  However these references adopted a particular
choice of gauge, transverse -- traceless gauge, which simplifies the canonical analysis but is not so natural from the point of view of AdS/CFT.  So we
will revisit these issues, in a gauge which is designed to make the bulk -- boundary correspondence as simple as possible.\footnote{From a technical point of view, compared to the work of
ADM, the main additional complication we face is that we will eventually impose a condition (\ref{Nz}) which fixes one of the components of the shift
vector.} 

As a prototype example we consider Einstein gravity coupled to a scalar field in an asymptotically anti-de Sitter space with dimension $D = d+1 > 3$.
\be
\label{scalarGR}
S = \int d^{d+1}x \, \sqrt{-G} \left(-{1 \over 2} G^{MN} \partial_M \phi \partial_N \phi + {1 \over 2 \kappa^2}(R - \Lambda)\right)
\ee
Notation: $x^M = (x^\mu,z)$ are bulk spacetime coordinates where $M=0,\ldots,d$.  $z$ is a radial coordinate in AdS, with $z \rightarrow 0$ at the boundary, and $x^\mu$, $\mu = 0,\ldots,d-1$ are
coordinates in the CFT.  We'll denote bulk spatial coordinates
including the radial direction by $x^i$, $i = 1,\ldots,d$, and bulk spatial coordinates excluding $z$ by $x^\hi$, $\hi = 1,\ldots,d-1$.
The cosmological constant is related to the anti-de Sitter radius by $\Lambda = - d(d-1)/R^2$.  We'll set $2 \kappa^2 = 1$ from now on.

First let's discuss our procedure for fixing bulk diffeomorphisms.  Similar issues arise for  gauge symmetries, as discussed in \cite{Heemskerk:2012mq,Kabat:2012hp,Kabat:2012av} and appendix \ref{appendix:gauge}.  In principle in the bulk we are free to make whatever choice of gauge we like.  But given the existence of a
boundary CFT, there is a preferred choice of coordinates in the bulk which makes the bulk -- boundary correspondence as simple as possible.
The following construction was first used to discuss bulk observables in AdS/CFT by Heemskerk \cite{Heemskerk:2012mq}.  The boundary has a preferred set of Minkowski coordinates $x^\mu$.  We can extend these
coordinates into the bulk by sending in geodesics perpendicular to the boundary.  Points along a given geodesic are then labeled by $x^\mu$ and the proper
distance $s$ measured from the boundary.  This proper distance diverges, of course, so to make it well-defined we introduce an IR cutoff in AdS and place the boundary at position $z' \rightarrow 0$.  Rather than work with proper distance directly, we define our radial coordinate $z$ in AdS by setting
\be
z = z' \exp(s / R)
\ee
Here $R$ is the AdS radius and we have in mind that the regulator can be removed by taking the limit $z' \rightarrow 0$, $s \rightarrow \infty$ with $z$ fixed.

From the geodesic equation ${d^2 x^A \over d\tau^2} + \Gamma^A_{BC} {dx^A \over d\tau} {dx^B \over d\tau} = 0$ the requirement that $x^\mu$ remain constant
along these radially-directed geodesics amounts to a restriction on the bulk Christoffel connection, namely that
\be
\label{GammaCondition}
\Gamma^\mu_{zz} = 0
\ee
This condition can be solved by putting the bulk metric in the form
\be
\label{FGgauge}
ds^2 = g_{\mu\nu} dx^\mu dx^\nu + {R^2 \over z^2} dz^2
\ee
These are known as Fefferman - Graham coordinates \cite{FG}, and were also used in \cite{Heemskerk:2012mq,Kabat:2012hp} to make the bulk -- boundary dictionary as simple as possible.
We will refer to this construction as fixing holographic gauge.

The next step is to understand how we impose holographic gauge on the theory (\ref{scalarGR}), and what the resulting commutation relations are.\footnote{Although we refer to commutation relations, our analysis will be purely classical, i.e.\ at the level of Poisson brackets.  Also we'll stop short of
determining the full set of commutators, since our goal is really just to show that matter fields have the usual equal-time commutators even in the presence of
gravity.}  To this end we consider the theory in a general gauge and introduce an ADM decomposition of the metric \cite{Arnowitt:1962hi}
\bea
&& G_{AB} = \left(\begin{array}{cc}
-N^2 + g_{ij} N^i N^j & g_{ij} N^j \\
g_{ij} N^j & g_{ij}
\end{array}\right) \\[15pt]
&& G^{AB} = \left(\begin{array}{cc}
-1/N^2 & N^i / N^2 \\
N^i / N^2 & g^{ij} - N^i N^j / N^2
\end{array}\right)
\eea
with $\sqrt{-G} = N \sqrt{g}$.
Spatial indices $i,j$ are raised and lowered with the metric $g_{ij}$.  Following ADM the theory can be put in first-order form, with Lagrangian
\be
\label{FirstOrder}
{\cal L} = \pi^{ij} \partial_t g_{ij} + \pi_\phi \partial_t \phi - N {\cal H} - N^i {\cal P}_i
\ee
The lapse $N$ and shift $N^i$ are Lagrange multiplier fields which enforce the Hamiltonian and momentum constraints
${\cal H} = {\cal P}_i = 0$.
\bea
\nonumber
\label{HamiltonianConstraint}
{\cal H} & = & {1 \over \sqrt{g}} \left(\pi^{ij} \pi_{ij} - {1 \over d - 1} \pi^2\right) - \sqrt{g} R(g) + {1 \over 2 \sqrt{g}} \pi_\phi^2 +
{1 \over 2} \sqrt{g} g^{ij} \partial_i \phi \partial_j \phi \\
& = & {1 \over \sqrt{g}} \left(\pi^{ij} \pi_{ij} - {1 \over d - 1} \pi^2\right) - \sqrt{g} R(g) + N^2 \sqrt{g} T^{00} \\[10pt]
\label{MomentumConstraint}
{\cal P}_i & = & - 2 \nabla_j \pi_i{}^j + \pi_\phi \partial_i \phi \\
& = & - 2 \nabla_j \pi_i{}^j - N \sqrt{g} T^0{}_i
\eea
Here $\nabla_i$ is the spatial covariant derivative (compatible with the metric $g_{ij}$), $R(g)$ is the curvature scalar constructed from $g_{ij}$, and $\pi \equiv g^{ij} \pi_{ij}$.  We've written the constraints both explicitly for a real scalar field, and more generally in terms of the matter stress tensor $T^{AB}$.  Along with
these constraints we impose four conditions to fix holographic gauge.  The appropriate gauge conditions turn out to be
\bea
\label{gf1}
g_{\hi z} & = & 0 \qquad \hi = 1,\ldots,d-1\\
\label{gf2}
g_{zz} & = & {R^2 \over z^2} \\
\label{gf3}
\pi_{zz} & = & {1 \over d - 1} g_{zz} \pi
\eea
The conditions (\ref{gf1}), (\ref{gf2}) aren't surprising; they put the spatial metric $g_{ij}$ in the Fefferman - Graham form (\ref{FGgauge}).  The condition
(\ref{gf3}) is less obvious.  It can be given several interpretations.
\begin{enumerate}
\item
In the ADM decomposition we have\footnote{See equation (3.9b) in \cite{Arnowitt:1962hi}.}
\be
\Gamma^0_{ij} = {1 \over N \sqrt{g}} \left(\pi_{ij} - {1 \over d-1} g_{ij} \pi\right)
\ee
Thus the condition (\ref{gf3}) sets $\Gamma^0_{zz} = 0$, which is the time component of the condition (\ref{GammaCondition}) on the bulk
Christoffel connection.
\item
In the ADM formalism the extrinsic curvature of the equal-time hypersurfaces is given by\footnote{See section 3.3 and footnote 3 in \cite{Arnowitt:1962hi}.}
\be
K_{ij} = - {1 \over \sqrt{g}} \left(\pi_{ij} - {1 \over d - 1} g_{ij} \pi\right)
\ee
Thus the condition (\ref{gf3}) sets $K_{zz} = 0$.
\item
As we will see below, it leads to the condition $N_z = 0$ or equivalently $G_{0z} = 0$.  It therefore puts the full metric (not just the spatial components)
in the Fefferman - Graham form (\ref{FGgauge}).
\end{enumerate}

At this stage it's useful to count degrees of freedom.  The metric $g_{ij}$ and its conjugate momentum $\pi^{ij}$ each have $d(d+1)/2$ independent components, for a total of $d(d+1)$ phase space degrees of freedom.  The Hamiltonian and momentum constraints eliminate $d+1$ degrees of freedom,
while the gauge conditions eliminate another $d+1$.  This leaves $d(d-1) - 2$ phase space degrees of freedom, or ${d(d - 1) \over 2} - 1$ configuration space
degrees of freedom, as expected for a massless spin-2 particle.\footnote{A symmetric traceless tensor under the $SO(d-1)$ little group.}

However we still have to fix the Lagrange multiplier fields $N$, $N^i$.  We do this by using equations of motion and requiring that the gauge-fixing
conditions (\ref{gf1}) -- (\ref{gf3}) are preserved under time evolution.\footnote{For the analogous procedure in transverse - traceless gravity see section
4.5 of \cite{Arnowitt:1962hi}.  The analogous procedure for Maxwell theory in axial gauge can be found on p.\ 80 of \cite{Hanson:1976cn}.}  For example
the equation of motion for the metric is\footnote{See equation (3.15a) in \cite{Arnowitt:1962hi}.  Although ADM consider pure gravity, this equation of motion
should remain the same in the presence of matter.}
\be
\partial_t g_{ij} = {2 N \over \sqrt{g}} \left(\pi_{ij} - {1 \over d-1} g_{ij} \pi\right) + \nabla_i N_j + \nabla_j N_i
\ee
This means
\be
\partial_t g_{zz} = {2 N \over \sqrt{g}} \left(\pi_{zz} - {1 \over d - 1} g_{zz} \pi\right) + 2 \nabla_z N_z
\ee
The left hand side vanishes by (\ref{gf2}), while the quantity in parenthesis vanishes by (\ref{gf3}), so with suitable boundary conditions as $z \rightarrow 0$
we are forced to set
\be
\label{Nz}
N_z = 0\,.
\ee
As promised, this condition on the shift puts the full spacetime metric in Fefferman - Graham form.  Then the $g_{\hi z}$ equation of motion reduces to
\be
\partial_t g_{\hi z} = {2 N \over \sqrt{g}} \pi_{\hi z} + \nabla_z N_\hi
\ee
The left hand side vanishes by (\ref{gf1}), so this becomes an equation we can solve to determine $N_\hi$.  Finally we need to impose the condition
that (\ref{gf3}) is preserved under time evolution.
\be
\partial_t \left(\pi_{zz} - {1 \over d - 1} g_{zz} \pi\right) = 0
\ee
This is a fairly complicated equation, but in principle it determines $N$ in terms of other degrees of freedom.\footnote{For $\partial_t \pi_{ij}$ in pure gravity
see (3.15b) in \cite{Arnowitt:1962hi}, but note that this equation is modified in the presence of matter.}

Now let's see about imposing our constraints and gauge-fixing conditions on the theory (\ref{FirstOrder}).  To do this we break the metric up into its
components $g_{\hi\hj}$,
$g_{\hi z}$ and $g_{zz}$.  Also it's useful to decompose
$g_{\hi\hj}$ into a conformal factor and a metric with unit determinant.  We do this by setting\footnote{This decomposition
was introduced by Dirac \cite{Dirac:1958jc}.  See \cite{Hanson:1976cn} p.\ 122.}
\bea
g_{\hi\hj} & = & e^{\sigma} \tilde{g}_{\hi\hj} \qquad \hbox{\rm with $\det \tilde{g}_{\hi\hj} = 1$} \\[5pt]
\pi^{\hi\hj} & = & e^{-\sigma} \left(\tilde{\pi}^{\hi\hj} + {1 \over d-1} \tilde{g}^{\hi\hj} \pi_\sigma\right) \qquad \hbox{\rm with $\tilde{g}_{\hi\hj} \tilde{\pi}^{\hi\hj} = 0$}
\eea
The condition $\det \tilde{g} = 1$ implies that $\tilde{g}^{\hi\hj} \partial_t \tilde{g}_{\hi\hj} = 0$.  So this is an orthogonal decomposition,
and the Lagrangian (\ref{FirstOrder}) becomes\footnote{For the corresponding Poisson brackets see (7.98) -- (7.101) in \cite{Hanson:1976cn}.}
\be
\label{FirstOrderTilde}
{\cal L} = \pi_\sigma \partial_t \sigma + \tilde{\pi}^{\hi\hj} \partial_t \tilde{g}_{\hi\hj} + 2 \pi^{\hi z} \partial_t g_{\hi z} + \pi^{zz} \partial_t g_{zz}
+ \pi_\phi \partial_t \phi - N {\cal H} - N^i {\cal P}_i
\ee
Our goal is to solve the constraints and gauge-fixing conditions for $\sigma$, $g_{\hi z}$, $g_{zz}$ and their
conjugate momenta $\pi_\sigma$, $\pi^{\hi z}$, $\pi^{zz}$.  This will leave a theory in which $\tilde{g}_{\hi\hj}$, $\tilde{\pi}^{\hi\hj}$, $\phi$, $\pi_\phi$
are the dynamical variables.  Now to work.
\begin{enumerate}
\item
The conditions (\ref{gf1}), (\ref{gf2}) can be directly imposed on the metric, they simply set $g_{\hi z} = 0$ and $g_{zz} = R^2 / z^2$.  The Lagrangian
reduces to
\be
\label{FirstOrderTilde2}
{\cal L} = \pi_\sigma \partial_t \sigma + \tilde{\pi}^{\hi\hj} \partial_t \tilde{g}_{\hi\hj} + \pi_\phi \partial_t \phi - N {\cal H} - N^i {\cal P}_i
\ee
Note that as a consequence of our gauge choice $\pi^{\hi z}$ and $\pi^{zz}$ drop out of the canonical $p \dot{q}$ part of the Lagrangian.
\item
The condition (\ref{gf3}) can be rewritten as
\be
\pi_{zz} = {1 \over d - 2} g_{zz} \pi_\sigma
\ee
This eliminates $\pi_{zz}$ as a dynamical variable, since it's proportional to the momentum conjugate to $\sigma$.
\item
The momentum constraint ${\cal P}_\hi = 0$ sets
\be
-2 \nabla_\hj \pi_\hi{}^\hj - 2 \nabla_z \pi_\hi{}^z + \pi_\phi \partial_\hi \phi = 0
\ee
which can be solved to determine $\pi^{\hi z}$.
\item
Likewise the constraint ${\cal P}_z = 0$, which now reads
\be
\label{Pz}
-2 \nabla_\hi \pi_z{}^\hi - {2 \over d - 2} \nabla_z \pi_\sigma + \pi_\phi \partial_z \phi = 0
\ee
can be solved to determine $\pi_\sigma$.
\item
Finally the Hamiltonian constraint (\ref{HamiltonianConstraint}), namely
\be
\label{H}
{1 \over \sqrt{g}} \left(\pi^{ij} \pi_{ij} - {1 \over d - 1} \pi^2\right) - \sqrt{g} R(g) + {1 \over 2 \sqrt{g}} \pi_\phi^2 +
{1 \over 2} \sqrt{g} g^{ij} \partial_i \phi \partial_j \phi = 0
\ee
can in principle be solved to determine $\sigma$.
\end{enumerate}

This completes our goal of eliminating all constraints and reducing the theory to the independent dynamical variables $\tilde{g}_{\hi\hj}$, $\tilde{\pi}^{\hi\hj}$, $\phi$, $\pi_\phi$.
However we still need to determine the commutators\footnote{more precisely Poisson brackets} for these physical degrees of freedom.  To do this
in principle
we should take the solutions for $\pi_\sigma$ and $\sigma$, plug into the Lagrangian (\ref{FirstOrderTilde2}), and read off the brackets from the resulting
symplectic form \cite{Faddeev:1988qp}.  This is a complicated procedure, which we will not attempt to carry out in detail.  However we do want to address the question of whether matter
fields still obey standard brackets when coupled to gravity -- in particular whether matter fields still commute with each other at spacelike separation -- since
this will form the basis of our CFT construction of local bulk observables.

The issue we face is the following.  Suppose we solve the system of constraints to determine $\sigma$.  We will get a (spatially non-local) expression of the
form
\be
\label{sigma}
\sigma = \sigma[\phi,\pi_\phi,\tilde{g}_{\hi\hj},\tilde{\pi}^{\hi\hj}]
\ee
When we take the time derivative of (\ref{sigma}) and plug the resulting expression for $\partial_t \sigma$ into (\ref{FirstOrderTilde2}), it would seem that time derivatives of $\phi$ and $\pi_\phi$ could
appear, which would modify the matter brackets.

Studying this in more detail, it becomes apparent that matter brackets are not modified by coupling to gravity.\footnote{We are grateful to Stanley Deser for reassuring us on this point.}  The reason is that matter only appears in the constraints through its conserved stress tensor.  More precisely it enters the Hamiltonian constraint (\ref{HamiltonianConstraint}) in the combination $N^2 \sqrt{g} T^{00}$, and it enters the momentum constraint (\ref{MomentumConstraint}) in the combination
$N \sqrt{g} T^0{}_i$.  It turns out these are exactly the combinations where, with the help of stress tensor conservation, all matter time derivatives can be eliminated
from $\partial_t \sigma$.

Let's see how this works in detail.  The Hamiltonian constraint
(\ref{HamiltonianConstraint}) depends on matter only through the combination $N^2 \sqrt{g} T^{00}$.  When we take a time derivative of this combination we generate the expression
\bea
\label{partialtT00}
\partial_t \left(N^2 \sqrt{g} T^{00}\right) & = & \left(\partial_t N \right) N \sqrt{g} T^{00} + N \partial_t \left(N \sqrt{g} T^{00}\right) \\
\nonumber
& = & \left(\partial_t N \right) N \sqrt{g} T^{00} + N \left(-\partial_i \left(N \sqrt{g} T^{i0}\right) - \Gamma^0_{AB} N \sqrt{g} T^{AB}\right)
\eea
where in the second line we used stress tensor conservation, $\nabla_A T^{A0} = 0$.  So far this still looks dangerous: $N$ is implicitly
a function of the matter fields, and the Christoffel symbols also have matter time derivatives hidden inside them.  Fortunately in the ADM decomposition of the metric the $\Gamma^0_{AB}$ Christoffel symbols are\footnote{The first two lines are a straightforward
calculation in ADM variables.  The third line follows from (3.9b) in \cite{Arnowitt:1962hi}.}
\bea
\label{Christoffel1}
\Gamma^{0}_{00} & = & {1 \over N} \partial_t N + {1 \over 2 N^2} \partial_t g_{ij} N^i N^j + {1 \over 2 N^2} N^i \partial_i \left(N^2 - N_k N^k\right) \\
\label{Christoffel2}
\Gamma^{0}_{0i} & = & {1 \over N} \partial_i N - {1 \over 2 N^2} \partial_i (N_k N^k) + {1 \over 2 N^2} N^j \left(\partial_t g_{ij} + \partial_i N_j - \partial_j N_i\right) \\
\label{Christoffel3}
\Gamma^{0}_{ij} & = & {1 \over N \sqrt{g}} \left(\pi_{ij} - {1 \over d - 1} g_{ij} \pi\right)
\eea
Using this in the second line of (\ref{partialtT00}), one finds that all terms involving $\partial_t N$ cancel.  So we're left with an expression for $\partial_t \left(N^2 \sqrt{g} T^{00}\right)$
that involves time derivatives
of the spatial metric $\partial_t g_{ij}$.  But there are no time derivatives of matter fields, nor are there time derivatives of $\pi^{ij}$.

Likewise in the momentum constraint (\ref{MomentumConstraint})
matter only enters through $N \sqrt{g} T^0{}_i$.  When we take a time derivative of this combination, due to the conservation equation $\nabla_A T^A{}_i = 0$
we get
\be
\label{partialtT0i}
\partial_t \left(N \sqrt{g} T^0{}_i\right) = - \partial_j \left(N \sqrt{g} T^j{}_i\right) + \Gamma^A_{Bi} N \sqrt{g} T^B{}_A
\ee
Now the relevant Christoffel symbols are, besides (\ref{Christoffel2}) and (\ref{Christoffel3}),
\bea
\nonumber
&&\Gamma^{i}_{0j} = - {1 \over N} N^i \partial_j N + {1 \over 2 N^2} N^i \partial_j (N_k N^k) + {1 \over 2} \left(g^{ik} - {N^i N^k \over N^2}\right) \left(\partial_t g_{jk} + \partial_j N_k - \partial_k N_j\right) \\
&&\Gamma^{i}_{jk} = \gamma^{i}_{jk} - {1 \over N \sqrt{g}} N^i \left(\pi_{ij} - {1 \over d - 1} g_{ij} \pi\right)
\eea
where $\gamma^{i}_{jk}$ is the connection constructed from the spatial metric $g_{ij}$.  So we're left with an expression for $\partial_t \left(N \sqrt{g} T^0{}_i\right)$ that involves time derivatives of the spatial metric $\partial_t g_{ij}$.  But again there are no time derivatives of matter fields, nor are there time derivatives of
$\pi^{ij}$.

In this way stress tensor conservation has
allowed us to eliminate matter time derivatives from our formula for $\partial_t \sigma$.  Moreover, as we have seen, the coupling to matter generates no time derivatives
of $\pi^{ij}$.\footnote{In general we do not expect that solving the constraints will generate terms involving $\partial_t \pi^{ij}$, since the Hamiltonian
formalism begins from an action that only involves first-order time derivatives of the metric.  We will assume this in the sequel.}  We still have
$\partial_t g_{ij}$ in our formulas, but note that
\bea
&& \partial_t g_{\hi\hj} = e^\sigma \partial_t \sigma \tilde{g}_{\hi\hj} + e^\sigma \partial_t \tilde{g}_{\hi\hj} \\
\nonumber
&& \partial_t g_{\hi z} = \partial_t g_{zz} = 0
\eea
This means the equation one obtains by taking a time derivative of (\ref{sigma}) can be solved to express $\partial_t \sigma$ in terms of the dynamical variables
$\tilde{g}_{\hi\hj}$, $\tilde{\pi}^{\hi\hj}$, $\phi$, $\pi_\phi$, but in a way
that does not involve time derivatives of matter fields or $\tilde{\pi}^{\hi\hj}$.  Substituting $\partial_t \sigma$ in (\ref{FirstOrderTilde2}), the symplectic form for matter retains its canonical form.  We show this in detail in appendix \ref{appendix:symplectic}.
This conclusion seems to hold quite generally, and applies to any theory in which the constraints (\ref{HamiltonianConstraint}), (\ref{MomentumConstraint}) only depend on matter through a conserved stress tensor.  {\em In such a theory, provided the gauge fixing conditions (\ref{gf1}) -- (\ref{gf3}) do not involve matter degrees of freedom, matter fields will
obey their usual canonical brackets even when coupled to gravity.}

\section{Gauge fields\label{sect:gauge}}
In this section we consider fields with spin one. While we are actually interested in the massless case (gauge fields), we will keep the discussion general.

Consider the 3-point function of two scalars and one field
strength $F_{\mu \nu}=\partial_{\mu}j_{\nu}-\partial_{\nu}j_{\mu}$, built from a current $j_{\mu}$ of dimension $\Delta$.  If the current is conserved then
$\Delta=d-1$, otherwise $\Delta > d-1$.
\begin{eqnarray}
& &\langle F_{\mu \nu}(x){\cal O}_{1}(y_1){\cal O}_{2}(y_2)\rangle  \sim   \frac{\left[(x-y_1)_{\mu}(x-y_2
)_{\nu}- \mu \leftrightarrow \nu \right]}{(y_1-y_2)^{\Delta_{1}+\Delta_{2}-\Delta+1}(x-y_1)^{\Delta+\Delta_1-\Delta_2+1}(y_2-x)^{\Delta+\Delta_2-\Delta_1+1}}
\nonumber \\
\label{f3point}
\end{eqnarray}
This can be written as a derivative operator acting on a scalar three-point function,
\begin{eqnarray}
& &\langle F_{\mu \nu}(x){\cal O}_{1}(y_1){\cal O}_{2}(y_2)\rangle  \sim \nonumber\\
& & \left[ (y_2-x)_{\mu}\frac{\partial}{\partial {x_\nu}}-\mu \leftrightarrow \nu \right] \frac{1}{(y_1-y_2)^{\Delta_{1}+\Delta_{2}-\Delta+1}(x-y_1)^{\Delta+\Delta_1-\Delta_2-1}(y_2-x)^{\Delta+\Delta_2-\Delta_1+1}}
\nonumber \\
\label{f3point1}
\end{eqnarray}
where the function the operator is acting on has the form of a  three-point function of three primary scalars of dimension $\Delta, \Delta_1, \Delta_2+1$ respectively. If we smear the scalar operator at the point $y_1$ into the bulk we will get the same derivative operator acting on the known result for a mixed bulk -- boundary scalar three-point function.\footnote{For a summary of the scalar case see appendix \ref{appendix:scalar}.} This has singularities at bulk spacelike separation which can be canceled, provided the current is not conserved, by adding smeared higher-dimension primary scalar operators to the definition of the bulk scalar. These operators can be constructed in $1/N$ perturbation theory as double-trace operators built from $j_{\mu}$, ${\cal O}_{2}$ and derivatives.  These higher-dimension primary scalars will also cancel the unwanted singularities in a three-point function with the current and another boundary primary scalar \cite{Kabat:2012av}.

However when $\Delta=d-1$, that is when $j_{\mu}$ is a conserved current, one cannot build a higher-dimension primary scalar out of the conserved current and another primary scalar. Even if one could, it wouldn't help: due to the Ward identity, the three-point function of a conserved current and two primary scalars is only non-zero
when the scalar operators have the same dimension.\footnote{See (33) in \cite{Kabat:2012av}.}

Fortunately, it turns out that for any current it is possible to construct from  $j_{\mu}$, ${\cal O}_{2}$ and derivatives a tower of non-primary scalar operators, which have correlation functions with $F_{\mu \nu}$ and ${\cal O}_{2}$ that take the form (\ref{f3point}) but with increasing $\Delta_{1}$. This is not a possibility we need to make
use of for
a non-conserved current.  But these non-primary operators will allow us to implement an appropriate notion of bulk micro-causality in the conserved current case.

Let us still be general and consider any current, conserved or not, and construct these non-primary scalars. At leading order in $1/N$ note that the most general scalar operator made out of $j_{\mu}$ (but not $\partial_\mu j^\mu$) and  ${\cal O}_{2}$, together with derivative operators, which transforms under dilations with naive
dimension $\Delta_{n}=\Delta_2+\Delta+2n+1$, is a sum of operators ${\cal A}^{mlk}$ given by
\begin{equation}
{\cal A}^{mlk}=\partial_{\mu_{1}\cdots \mu_{m}}\nabla^{2l}j_{\nu}\partial^{\mu_{1}\cdots \mu_{m}}
\nabla^{2k}\partial^{\nu}{\cal O}_{2}
\end{equation}
We want to find an operator ${\cal A}_{n}$ whose correlator has the form (\ref{f3point}) with $\Delta_1 \rightarrow \Delta_{n}$, namely
\be
\langle F_{\mu \nu}(x){\cal A}_{n}(y_1){\cal O}_{2}(y_2)\rangle = \frac{\left[(y_1-x)_{\mu}(y_2-x)_{\nu}- \mu \leftrightarrow \nu \right](x-y_2)^{2n}}{(y_1-y_2)^{2\Delta_{2}+2+2n}(x-y_1)^{2\Delta+2+2n}}
\label{3fn}
\ee
To construct such an operator we write the right hand side of (\ref{3fn}) as 
\be
\left[(x-y_1)_{\mu}(x-y_2)_{\nu}- \mu \leftrightarrow \nu \right]\sum_{m+l+k=n}d_{mlk}\frac{[(y_1-y_2)_{\alpha}(x-y_1)^{\alpha}]^{m}}{(x-y_1)^{2\Delta+2+2m+2l}(y_1-y_2)^{2\Delta_2 +2+2m+2k}}
\label{3fmlk}
\ee
where
\be
d_{mlk}=2^{m}\left({m+l+k \atop m}\right)\left({l+k \atop l}\right).
\ee
In appendix \ref{appendix:operators} we show that we can construct operators ${\cal V}^{mlk}$ built from linear combinations of the ${\cal A}^{mlk}$, where
$\langle F_{\mu \nu}(x){\cal V}^{mlk}(y_1){\cal O}_{2}(y_2)\rangle$ exactly matches the corresponding term in the expansion written in (\ref{3fmlk}). Then we can write
\be
\label{A_n}
{\cal A}_{n}=\frac{1}{N} \sum_{m+l+k=n} d_{mlk}{\cal V}^{mlk}
\ee
This is true whether or not the current is conserved, since we have not used the operator $\partial^{\mu} j_{\mu}$ anywhere. For a conserved current the same
formulas hold, with the following changes of notation.  In the original three-point function (\ref{f3point}) we replace ${\cal O}_1 \rightarrow {\cal O}$ and take ${\cal O}_2$
to be its complex conjugate, ${\cal O}_2 \rightarrow \bar{\cal O}$.  We take the dimension of the current $\Delta = d - 1$, and in the expressions for ${\cal V}_{mlk}$
we replace ${\cal O}_2 \rightarrow {\cal O}$.

Now using the result from smearing a scalar operator inside a scalar three-point function \cite{Kabat:2011rz,Kabat:2012av},
we see that we can define a local bulk scalar field interacting with a bulk gauge field by setting
\be
\phi(z,y_1)=\int K_{\Delta_{1}}{\cal O}+\sum_{n=0}^{\infty} a_{n} \int K_{\Delta_{n}} {\cal A}_{n}
\label{bsg}
\ee
Here $K_{\Delta}(z,y|y')$ is the scalar smearing function for a dimension $\Delta$ primary scalar.
With appropriately chosen $a_n$, all the unwanted space-like singularities can be canceled in a three-point function of this operator with a boundary
field strength $F_{\mu \nu}$ and a boundary scalar $\bar{\cal O}$.

Note that if we had been considering a scalar field interacting with a massive vector field in the bulk, we would not need to consider the non-primary
operators ${\cal A}_{n}$.  Rather we would cancel the unwanted singularities using the higher-dimension primary scalars $\sim \partial_\mu j^\mu \, {\cal O}_2 + j^\mu \partial_\mu {\cal O}_2$ that one can build from a non-conserved current $j_\mu$ 
and ${\cal O}_{2}$.  This procedure would allow us to build a bulk scalar which is local in correlation functions involving $j_\mu$ \cite{Kabat:2012av}.
For a massive vector, we could use non-primary scalars if we were only interested in
achieving locality in correlators involving the boundary field strength
$F = dj$.  But since locality would then be violated in correlators involving the current $j$
itself, this procedure is not physically sensible.

\subsection{Special conformal transformations}

In this section we show that bulk micro-causality implies that our bulk observables have the correct transformation properties under AdS isometries.
The issue is that the operators ${\cal A}_{n}$ we have constructed are not primary scalars, but in (\ref{bsg}) they are smeared using the usual scalar
smearing function.  This means the bulk operator defined in (\ref{bsg}) will not transform like an ordinary scalar field under AdS isometries.  Instead, as we
will show, it obeys the correct transformation rule for a charged scalar field in holographic gauge.

We start with a non-primary scalar operator ${\cal A}_{n}$, whose three-point function with $F_{\mu \nu}$ and a primary scalar of dimension $\Delta_2$ is
\begin{eqnarray}
& &\langle F_{\mu \nu}(x){\cal A}_{n}(y_1){\cal O}_{2}(y_2)\rangle  \sim \nonumber\\
& &  \frac{1}{(y_1-y_2)^{\Delta_{n}+\Delta_{2}-\Delta+1}(y_1-x)^{\Delta+\Delta_n-\Delta_2+1}(y_2-x)^{\Delta+\Delta_2-\Delta_n+1}}\left[(y_1-x)_{\mu}(y_2-x)_{\nu}- \mu \leftrightarrow \nu \right]
\nonumber \\
\label{faphi}
\end{eqnarray}
This restricts how ${\cal A}_n$ transforms under conformal transformation. Of course one possibility is that ${\cal A}_n$ is a primary scalar, but we will see that there is another possibility.
To see how ${\cal A}_n$ does behave, we transform both sides of the equality and ask how they can match.
Since ${\cal A}_{n}$ transforms as a scalar under rotations, and as a scalar with dimension $\Delta_{n}$ under dilations, we only need to see what happens under special conformal transformations.
Under a special conformal transformation with parameter $b_{\mu}$, to linear order in $b_{\mu}$ one has the transformation properties
\begin{eqnarray}
 F'_{\rho \mu}&=&F_{\rho \mu}(1-2(\Delta_j+1)(b\cdot x))-2b_{\mu}x^{\nu}F_{\rho \nu}+2b_{\rho}x^{\nu}F_{\mu \nu}\nonumber \\
&&+2x_{\mu}b^{\nu}F_{\rho \nu}-2x_{\rho}b^{\nu}F_{\rho \nu}+2(\Delta_{j}-1)(b_{\mu}j_{\rho}-b_{\rho}j_{\mu}) \nonumber\\[5pt]
 {\cal O}'_{2} &=&(1-2\Delta_{2}b\cdot x) {\cal O}_{2}
\label{ftrans}
\end{eqnarray}
Let us split the transformation of ${\cal A}_n$ into a piece $\delta_s{\cal A}_{n}$ which is the expected behavior if it was a primary scalar of dimension $\Delta_{n}$,
and an extra piece $\delta_e{\cal A}_{n}$.
Since the right hand side of (\ref{faphi}) transforms as if ${\cal A}_n$ were a primary scalar, there must be some cancellations on the left hand side to achieve this.
Under a special conformal transformation the left hand side of (\ref{faphi}) changes by
\begin{equation}
\langle\delta F_{\mu \nu} {\cal A}_n {\cal O}_{2}\rangle +\langle F_{\mu \nu} \delta{\cal A}_n {\cal O}_{2}\rangle +\langle F_{\mu \nu} {\cal A}_n \delta{\cal O}_{2}\rangle 
\end{equation}
Since ${\cal A}_{n}$ behaves like a primary scalar in a three-point function with $F_{\mu \nu}$ and ${\cal O}_{2}$ (but not with $j_{\mu}$), the only terms which are
sensitive to the fact that ${\cal A}_n$ is actually not a primary scalar are
\begin{equation}
\langle 2(\Delta_{j}-1)(b_{\nu}j_{\mu}-b_{\mu}j_{\nu}) {\cal A}_n {\cal O}_{2}\rangle + \langle F_{\mu \nu} \delta_e {\cal  A}_{n} {\cal O}_{2}\rangle 
\label{transeq}
\end{equation}
But again, since ${\cal A}_n$ has a three-point function like a primary scalar with $F_{\mu \nu}$ and ${\cal O}_{2}$, it must be the case that
\begin{equation}
\langle j_{\mu} {\cal A}_n {\cal O}_{2}\rangle = (\hbox{\rm primary scalar result}) +\partial_{\mu} B
\end{equation}
where the last term has a vanishing exterior derivative and drops out of the three-point function with $F$. 
This means the first term in (\ref{transeq}) has the form
\begin{equation}
(b_{\nu}\partial_{\mu}-b_{\mu}\partial_{\nu})B(x,y_1,y_2)
\label{gform}
\end{equation}
Now in order for (\ref{transeq})  to vanish, it must be the case that
$\delta_e {\cal A}_{n} $ is composed of terms of the form
\begin{equation}
 \partial_{\alpha_1 \cdots \alpha_{n_1}}(\nabla^{2n_2}b \cdot j) \partial^{\alpha_1 \cdots \alpha_{n_1}}\nabla^{2n_3}{\cal O}_{2},
 \label{vtrans}
\end{equation}
since only then will $\langle F_{\mu \nu} \delta_e {\cal  A}_{n} {\cal O}_{2}\rangle$ have the form (\ref{gform}).  That is, the vector index of the transformation parameter
$b_\rho$ must be contracted with the index on $j_\rho$, and not with one of the derivative operators.
This is verified by explicit computation for the two lowest-dimensions operators in appendix \ref{appendix:conformal}.

This means that under special conformal transformations the expression for the bulk field in (\ref{bsg}) transforms as
\be
\phi'(z',y'_1)=\phi(z,y_1)+\sum_{n=0}^{\infty} a_{n} \int K_{\Delta_{n}} \delta_e {\cal A}_{n}
\ee
Fortunately this is exactly the type of transformation that a charged bulk scalar field should have.
To see this recall how a charged bulk field behaves under special conformal transformation.\footnote{This was discussed in \cite{Kabat:2012av}.}
In a completely fixed gauge the degrees of freedom which are left are physical, but they may only look local in the chosen gauge. For example a charged
matter field in holographic gauge $\phi_{\rm phys}$ can be written in terms of the non-gauge-fixed variables as
\begin{equation}
\label{phi_phys}
\phi_{\rm phys}(z,y_1)=e^{\frac{1}{N}\int_{0}^{z} A_{z} dz}\phi(z,y_1)
\end{equation}
where we have attached a Wilson line running to the boundary, and $1/N$ is the charge of the field.  In holographic gauge $A_z = 0$ and the Wilson line
is invisible.  This expression makes it manifest that the matter degrees of freedom in holographic gauge are secretly non-local.

One can directly compute how special conformal transformations act on the right hand side of (\ref{phi_phys}).  Alternatively one can realize that since special conformal transformations do not preserve the gauge $A_z=0$, they must be combined with a compensating gauge transformation chosen to restore holographic gauge.  Only
the combined transformation is a symmetry of the gauge-fixed theory.  This tells us how operators in holographic gauge should behave under special conformal
transformations, namely
\begin{equation}
\phi'_{\rm phys}(z',y'_1)=e^{-\frac{i}{N}\lambda(z,y_1)}\phi_{\rm phys}(z,y_1)
\end{equation}
where the compensating gauge transformation is \cite{Kabat:2012hp}
\begin{equation}
\lambda=-\frac{1}{\rm vol(S^{d-1})}\int d^d x'' \theta (\sigma z'')2b\cdot j
\end{equation}
To leading order in $1/N$ 
\begin{equation}
\phi'_{\rm phys}(z',y'_1)=\phi_{\rm phys}(z,y_1)-\frac{i}{N}\lambda (z,y_1)\phi_{0}(z,y_1)
\end{equation}
where
$\phi_{0}=\int K(z,y_1,x''){\cal O}(x'')$ is the zeroth-order smeared operator, without any interaction corrections. The term $\lambda (z,y_1)\phi_{0}(z,y_1)$ is a bi-local smeared expression on the boundary involving the operators $b\cdot j(x_1){\cal O}(x_2)$, and hence should have a Taylor expansion around $b\cdot  j(y_1) {\cal O}(y_1)$ involving exactly the operators appearing in (\ref{vtrans}).

\subsection{Scalar commutator}

We have shown that, even when the current is conserved, one can construct the double-trace operators ${\cal A}_n$ given in (\ref{A_n}).  These operators
are scalars but are not primary.  They have the feature that, even though they are not primary scalars, they behave like a primary scalar when inserted in a
three-point function with $F_{\mu\nu}$ and another primary scalar.  That is, the correlation function (\ref{3fn}) has the same form as (\ref{f3point}).  Then given (\ref{f3point1}) one can define an interacting bulk scalar field by smearing the operators ${\cal A}_n$ as though they were primary scalars and adding them to the
zeroth-order definition of the bulk field with arbitrary coefficients as in (\ref{bsg}).  By choosing the coefficients $a_n$ appropriately, we can make the commutator between the bulk field
and the boundary primary scalar ${\cal O}_{2}$ or the boundary field strength $F_{\mu \nu}$ as small as we like at spacelike separation, at least inside the
three-point function.  This procedure is directly
analogous to the case of interacting scalar fields.

Note that this procedure only addresses the commutator in a three-point function $\langle \phi F \bar{\cal O} \rangle$ involving the bulk scalar, a boundary field strength $F_{\mu\nu}$, and another boundary
scalar.  So it does not guarantee a vanishing commutator between the bulk field and the boundary current.  Indeed we expect the commutator of the bulk scalar with
the boundary current to be non-zero at
spacelike separation due to the Gauss constraint.  But will now argue that the procedure does imply a vanishing commutator between the bulk field and the boundary scalar in a
three-point function with the current.  That is, we claim that for conserved currents, to leading order in $1/N$
\begin{equation}
\langle F_{\mu \nu}(x)[\phi(z,y_1), {\cal O}_{2}(y_2)]\rangle =0 \quad \Rightarrow \quad \langle j_{\mu}(x)[\phi(z,y_1), {\cal O}_{2}(y_2)]\rangle =0
\label{zerocom}
\end{equation}
The argument is as follows. 

For a three-point function involving the commutator and the current $\langle [\phi,\bar{\cal O}] j \rangle$ to be non-zero at leading order in the $1/N$ expansion the commutator must be proportional to an operator linear in the current. Then for the commutator to have a vanishing two-point function with $F_{\mu \nu}$, i.e.\ $\langle [\phi,\bar{\cal O}] F \rangle = 0$,
the commutator must be proportional to the divergence of the current. But for a
conserved current the divergence vanishes, and this implies the right hand side of (\ref{zerocom}). Thus while the addition of higher-dimension non-primary scalar operators
can cancel the spacelike commutator with another boundary scalar, it cannot cancel the non-vanishing commutator with the current.  This is, of course, required by
the bulk Gauss constraint.  Note that the same logic cannot be used to show that if the bulk scalar commutes with $F_{\mu \nu}$ it will also commute with $j_{\mu}$.
The reason is that a vanishing commutator with $F_{\mu\nu}$ allows a non-zero commutator with the current of the form
\be
[j_{\mu}(x),\phi(z,y_1)] \sim \partial^{x}_{\mu}(\int dx'' c(z,y_1,x,x''){\cal O}_{2}(x''))
\ee
 
\section{Gravity\label{sect:gravity}}
We now turn to bulk scalar fields interacting with gravity.  We will follow a similar route to the previous section, and will arrive at
similar conclusions, but instead of working with a conserved current we will work with the conserved energy-momentum tensor of the CFT.

The three-point function of the energy-momentum tensor and two primary scalars of dimension $\Delta$  is given by
\begin{eqnarray}
& &\langle T_{\mu \nu}(x){\cal O}(y_1){\cal O}(y_2)\rangle =\frac{c_{d,\Delta}}{(x-y_1)^{d-2}(x-y_2)^{d-2}(y_1-y_2)^{2\Delta-d+2}}\nonumber\\
& &\left[\left(\frac{(x-y_1)_{\mu}}{(x-y_1)^2}-\frac{(x-y_2)_{\mu}}{(x-y_2)^2}\right)\left(\frac{(x-y_1)_{\nu}}{(x-y_1)^2}-\frac{(x-y_2)_{\nu}}{(x-y_2)^2}\right)
-\frac{\eta_{\mu \nu}}{d}\left(\frac{(x-y_1)_{\rho}}{(x-y_1)^2}-\frac{(x-y_2)_{\rho}}{(x-y_2)^2}\right)^2\right] \nonumber\\
\label{3pointt}
\end{eqnarray}
This can be written as a second-order derivative operator with respect to $x$, and some functions of $(x-y_2)$, acting on
\begin{equation}
\frac{1}{(x-y_1)^{d-2}(x-y_2)^{d-2}(y_1-y_2)^{2\Delta-d+2}}\nonumber
\end{equation}
This expression is the three-point function of scalar primaries of dimension $d-2,\Delta,\Delta$.\footnote{We leave aside the issue of operators with such low dimensions.}
One can smear the operator ${\cal O}(y_1)$ to move it into the bulk.  Then one gets the same derivative operator acting on a smeared scalar three-point function, whose analytic structure we know. To make a local bulk scalar one would need a tower of operators of dimension $\Delta_{n}$, whose three-point function would resemble
(\ref{3pointt})  with
\begin{equation}
\frac{1}{(x-y_1)^{d-2}(x-y_2)^{d-2}(y_1-y_2)^{2\Delta-d+2}}\rightarrow \frac{1}{(x-y_1)^{d-2+\Delta_n -\Delta}(x-y_2)^{d-2+\Delta-\Delta_n}(y_1-y_2)^{\Delta+\Delta_n-d+2}}
\end{equation}
But such operators do not exist. In fact, due to the Ward identity, the three-point function of the energy-momentum tensor with two primary scalars can only be
non-zero if the scalars have the same dimension.  (The same issue arose with a conserved current in the previous section.)
The inability to construct such operators out of $T_{\mu \nu}$ and ${\cal O}$ can be traced to the absence of operators associated with the divergence of the energy-momentum tensor.  For a non-conserved spin-2 tensor there would be no such obstruction.  This breakdown of locality is desirable, since the Hamiltonian can
be written as a surface integral (a spatial integral of $T^{00}$), and the Hamiltonian cannot commute with any bulk operator that is not a constant of the motion.

Instead we wish to proceed as in the previous section, and see if we can make a bulk scalar which commutes at spacelike separation with another boundary
scalar when inserted in a three-point function.  For this we need to find a gravity operator analogous to $F_{\mu \nu}$.
It turns out the appropriate choice is the boundary Weyl tensor with all indices taken to have distinct values.\footnote{We work in dimension $d \geq 4$ so that this is possible.}  This is given by
\begin{equation}
C_{\alpha \beta \gamma \delta}=\partial_{\alpha}\partial_{\gamma}T_{\beta \delta}-\partial_{\alpha}\partial_{\delta}T_{\beta \gamma}-\partial_{\beta}\partial_{\gamma}T_{\alpha \delta}+\partial_{\beta}\partial_{\delta}T_{\alpha \gamma}
\end{equation}
For later use it is important that the expression for the Weyl tensor is invariant under
\begin{equation}
T_{\mu \nu} \rightarrow T_{\mu \nu} +\partial_{\mu} \epsilon _{\nu}+\partial_{\nu} \epsilon _{\mu}.
\end{equation}
The three-point function of $C_{\alpha \beta \gamma \delta}$ with two primary scalars of dimension $\Delta$ is
\begin{eqnarray}
&& \langle C_{\alpha \beta \gamma \delta}(x) {\cal O}(y_1){\cal O}(y_2)\rangle =\frac{-4dc_{d,\Delta}}{(x-y_1)^{d+2}(x-y_2)^{d+2}(y_1-y_2)^{2\Delta-d+2}}\nonumber\\
&& \quad \left[(x-y_1)_{\beta}(x-y_1)_{\delta}(x-y_2)_{\alpha}(x-y_2)_{\gamma}-(\gamma \leftrightarrow \delta)-(\beta \leftrightarrow \alpha)+(\gamma \leftrightarrow \delta \ \ \beta \leftrightarrow \alpha)\right]\nonumber\\
\label{css}
\end{eqnarray}
This can be written as 
\begin{eqnarray}
& &\langle C_{\alpha \beta \gamma \delta}(x) {\cal O}(y_1){\cal O}(y_2)\rangle \sim {\cal L}_{g}
\frac{1}{(x-y_1)^{d-2}(x-y_2)^{d+2}(y_1-y_2)^{2\Delta-d+2}}\nonumber\\
& &{\cal L}_{g}=\left[(x-y_2)_{\alpha}(x-y_2)_{\gamma}\partial_{x_\beta}\partial_{x_\delta}-(\gamma \leftrightarrow \delta)-(\beta \leftrightarrow \alpha)+(\gamma \leftrightarrow \delta \ \ \beta \leftrightarrow \alpha)\right]\nonumber\\
\label{css1}
\end{eqnarray}
where the operator ${\cal L}_g$ is acting on the three-point function of primary scalars of dimensions  ($d,\Delta,\Delta+2$). If we try to promote ${\cal O}(y_1)$ to a bulk operator by smearing it we will obtain a three-point function that has singularities at bulk spacelike separation.  To cancel these singularities we will need to add
appropriately smeared higher-dimension scalar operators. Thus we need to find a tower of scalar operators ${\cal T}_n$ which transform under dilations with increasing dimensions
$\Delta_{n}=\Delta+d+2+2n$, and whose three-point function with $C_{\alpha \beta \gamma \delta}(x)$ and ${\cal O}(y_2)$ matches (\ref{css1}) with
\begin{equation}
\frac{1}{(x-y_1)^{d-2}(x-y_2)^{d+2}(y_1-y_2)^{2\Delta-d+2}} \rightarrow \frac{1}{(x-y_1)^{d-2+\Delta_{n}-\Delta}(x-y_2)^{d+2+\Delta-\Delta_n}(y_1-y_2)^{\Delta+\Delta_n-d+2}}
\label{repcss}
\end{equation}
Note that for  $\Delta_{n}=\Delta+d+2+2n$ this becomes
\be
\frac{(x-y_2)^{2n}}{(x-y_1)^{2d+2n}(y_1-y_2)^{2\Delta+2n+4}} \
\ee
So we are looking for operators ${\cal T}_{n}$ which obey
\bea
&&\langle C_{\alpha \beta \gamma \delta}(x) {\cal T}_{n}(y_1){\cal O}(y_2)\rangle =\frac{(x-y_2)^{2n}}{(x-y_1)^{2d+2n+4}(y_1-y_2)^{2\Delta+2n+4}}\times\nonumber\\
&&\left[(x-y_1)_{\beta}(x-y_1)_{\delta}(x-y_2)_{\alpha}(x-y_2)_{\gamma}-(\gamma \leftrightarrow \delta)-(\beta \leftrightarrow \alpha)+(\gamma \leftrightarrow \delta \ \ \beta \leftrightarrow \alpha)\right]\nonumber\\
\label{csstran}
\eea
As in the vector case it is useful to write
\be
(x-y_2)^{2n}=\sum_{m+k+l=n}d_{mlk}(x-y_1)^{2k}(y_1-y_2)^{2l}[(y_1-y_2)_{\alpha}(x-y_1)^{\alpha}]^{m}
\ee
and look for operators ${\cal M}_{mlk}$ obeying
\bea
&&\langle C_{\alpha \beta \gamma \delta}(x) {\cal M}_{mlk}(y_1){\cal O}(y_2)\rangle =\frac{[(y_1-y_2)_{\alpha}(x-y_1)^{\alpha}]^{m}
}{(x-y_1)^{2d+2m+2l+4}(y_1-y_2)^{2\Delta+2m+2k+4}}\times\nonumber\\
&&\left[(x-y_1)_{\beta}(x-y_1)_{\delta}(x-y_2)_{\alpha}(x-y_2)_{\gamma}-(\gamma \leftrightarrow \delta)-(\beta \leftrightarrow \alpha)+(\gamma \leftrightarrow \delta \ \ \beta \leftrightarrow \alpha)\right]\nonumber\\
\eea
To leading order in $1/N$ these operators can be constructed starting from the most general scalar operator with the correct dimension
\begin{equation}
{\cal M}_{mlk}=\sum_{m+l+k=m'+l'+k'}b^{m'l'k'}_{mlk} \partial_{\mu_{1}\cdots \mu_{m'}}\nabla^{2l'}T_{\rho \nu}\partial^{\mu_{1}\cdots \mu_{m'}}
\nabla^{2k'}\partial^{\nu}\partial^{\rho}{\cal O}_{2}
\end{equation}
and solving for the coefficients $b$.  In appendix \ref{appendix:operators} we give an iterative construction of these coefficients.
The desired scalar non-primary operators are then
\be
{\cal T}_{n}=\sum_{m+l+k=n}d_{mlk} {\cal M}_{mlk}
\ee
For example when $n=0$ we have ${\cal T}_0 = T_{\rho \nu}\partial^{\nu}\partial^{\rho}{\cal O}$, and when
$n=1$ we find
\begin{equation}
b^{0,0,1}_{0,0,1}=\frac{C_1}{(2\Delta+2+d)(2\Delta +4)},\ \ b^{1,0,0}_{1,0,0}=\frac{C_1}{(2+d)(2d +4)},\ \ b^{0,1,0}_{0,1,0}=\frac{-C_1}{(2d+4)(2\Delta +4)}\ \
\end{equation}
where $C_1=\frac{1}{32d^2(d^2-1)\Delta(\Delta+1)}$.
Much like the gauge field case, the fact that we cannot construct a primary scalar from $T_{\mu\nu}$ and ${\cal O}_2$ is due to stress tensor conservation, $\partial_\mu T^{\mu \nu} = 0$.

Given these operators, we can define a bulk scalar field that has a micro-causal three-point function with a boundary primary scalar and the boundary Weyl tensor by
setting
\be
\phi(z,y_1)=\int K_{\Delta_{1}}{\cal O}+\sum_{n=0}^{\infty} b_{n} \int K_{\Delta_{n}} {\cal T}_{n}
\label{bstt}
\ee
The constants $b_n$ are chosen so that the unwanted space-like singularities in the three-point function are canceled.

\subsection{Special conformal transformation}
Since ${\cal T}_{n}$ is not a primary scalar, the field $\phi(z,y_1)$ defined in (\ref{bstt}) will not transform like a conventional bulk scalar field under AdS isometries.
Instead, as we will see, it has the correct transformation properties to represent a scalar field in holographic gauge.  Thus somewhat remarkably, just as in the gauge field
case, imposing micro-causality leads to bulk fields with the correct transformation properties.

To determine how $\phi(z,y_1)$ transforms, we first need to know how ${\cal T}_{n}$ transforms. Rather than constructing ${\cal T}_n$ explicitly and finding its transformation properties, we use an alternate route. 
We wish to determine the transformation of a scalar operator ${\cal T}_{n}$ whose 3-point function
\begin{equation}
\langle C_{\alpha \mu \beta \nu}(x){\cal T}_{n}(y_1){\cal O}(y_2)\rangle 
\end{equation}
obeys (\ref{csstran}). We follow the same logic as in the conserved current case, and look for the behavior under infinitesimal special conformal transformations which
would not be present if ${\cal T}_{n}$ was a primary scalar.

We start with the behavior of the energy momentum tensor under a special conformal transformation.  To first order in the parameter $b_{\mu}$
\begin{eqnarray}
T'_{\mu \nu}=T_{\mu \nu}(1-2d(b\cdot x))+2b^{\delta}x_{\nu}T_{\mu \delta}-2b_{\nu}x^{\delta}T_{\mu \delta}+2b^{\delta}x_{\mu}T_{\delta \nu}-2b_{\mu}x^{\delta}T_{\delta \nu}
\end{eqnarray}
The transformation of $C_{\alpha \mu \beta \nu}$ when all indices are distinct is
\begin{eqnarray}
& & C'_{\alpha \mu \beta \nu}=C_{\alpha \mu \beta \nu}(1-2(d+2)b\cdot x)+2b^{\delta}x_{\nu}C_{\alpha \mu \beta \delta}+2b^{\delta}x_{\mu}C_{\alpha \delta \beta \nu}+2b^{\delta}x_{\beta}C_{\alpha \mu \delta \nu}+2b^{\delta}x_{\alpha}C_{\delta \mu \beta \nu}\nonumber\\
& & -2x^{\delta}b_{\nu}C_{\alpha \mu \beta \delta}-2x^{\delta}b_{\mu}C_{\alpha \delta \beta \nu}-2x^{\delta}b_{\beta}C_{\alpha \mu \delta \nu}-2x^{\delta}b_{\alpha}C_{\delta \mu \beta \nu}\nonumber\\
& & -2d[(b_{\alpha}\partial_{\beta}+b_{\beta}\partial_{\alpha})T_{\mu \nu}+(b_{\mu}\partial_{\nu}+b_{\nu}\partial_{\mu})T_{\alpha \beta}-(b_{\alpha}\partial_{\nu}+b_{\nu}\partial_{\alpha})T_{\mu \beta}-(b_{\mu}\partial_{\beta}+b_{\beta}\partial_{\mu})T_{\alpha \nu}]\nonumber\\
\end{eqnarray}
Let us look at the transformation of the left hand side of (\ref{csstran}), namely
\begin{equation}
\langle \delta C_{\alpha \mu \beta \nu}(x) {\cal T}_n {\cal O}_{2}\rangle +\langle C_{\alpha \mu \beta \nu}(x) \delta{\cal T}_n {\cal O}_{2}\rangle +\langle  C_{\alpha \mu \beta \nu}(x) {\cal T}_n \delta{\cal O}_{2}\rangle 
\end{equation}
Let us also split the transformation of ${\cal T}_{n}$ into a piece $\delta_{s}{\cal T}_{n}$, which is the part that looks like the transformation of a primary scalar with dimension $\Delta_{n}$, and an extra piece $\delta_{\rm extra}{\cal T}_{n}$.
Since by assumption ${\cal T}_{n}$ obeys (\ref{css1}) and (\ref{repcss}), the only terms which differ from the case that ${\cal T}_n$ is actually a primary scalar are
\begin{eqnarray}
& & \langle [(b_{\alpha}\partial_{\beta}+b_{\beta}\partial_{\alpha})T_{\mu \nu}+(b_{\mu}\partial_{\nu}+b_{\nu}\partial_{\mu})T_{\alpha \beta}-(b_{\alpha}\partial_{\nu}+b_{\nu}\partial_{\alpha})T_{\mu \beta}-\nonumber\\
& &(b_{\mu}\partial_{\beta}+b_{\beta}\partial_{\mu})T_{\alpha \nu}](x){\cal T}_{n}(y_1){\cal O}(y_2)\rangle 
  +\langle C_{\alpha \mu \beta \nu}(x)\delta_{\rm extra}{\cal T}_{n}(y_1) {\cal O}(y_2)\rangle \nonumber\\
  \label{gextra}
\end{eqnarray}
However since $\langle C_{\alpha \mu \beta \nu}(x){\cal T}_{n}(y_1){\cal O}(y_2)\rangle $ transforms as if ${\cal T}_{n}$ is a scalar, then
\begin{equation}
\langle T_{\mu \nu} {\cal T}_{n} {\cal O}\rangle = \hbox{\rm scalar case} + \partial_{\mu}B_{\nu}+\partial_{\nu}B_{\mu}.
\end{equation}
This means that  for (\ref{gextra}) to vanish one  must have
\begin{eqnarray}
& &\langle C_{\alpha \mu \beta \nu}(x)\delta_{\rm extra}{\cal T}_{n}(y_1) {\cal O}(y_2)\rangle  \sim (b_{\alpha}\partial_{\beta}\partial_{\mu}-b_{\mu}\partial_{\beta}\partial_{\alpha})B_{\nu}+
(b_{\beta}\partial_{\alpha}\partial_{\nu}-b_{\nu}\partial_{\beta}\partial_{\alpha})B_{\mu}\nonumber\\
&& \quad + (b_{\nu}\partial_{\beta}\partial_{\mu}-b_{\beta}\partial_{\mu}\partial_{\nu})B_{\alpha}+
(b_{\mu}\partial_{\alpha}\partial_{\nu}-b_{\alpha}\partial_{\mu}\partial_{\nu})B_{\beta}.
\end{eqnarray}
This can only be achieved if $\delta_{\rm extra}{\cal T}_{n}$ is made out of terms of the form
\begin{equation}
 \partial_{\alpha_1 \cdots \alpha_{n_1}}(\nabla^{2n_2}b^{\delta} T_{\delta \rho})\partial^{\rho} \partial^{\alpha_1 \cdots \alpha_{n_1}}\nabla^{2n_3}{\cal O}_{2}
 \label{ttrans}
\end{equation}
No other types of contraction of $b_{\mu}$ will give the right result.
Explicit computations in appendix \ref{appendix:conformal} for ${\cal T}_{0}$ agree with this form.
Thus under special conformal transformation the bulk scalar (\ref{bstt}) transforms as
\be
\phi'(z',y'_1)=\phi(z,y_1)+\sum_{n=0}^{\infty} b_{n} \int K_{\Delta_{n}} \delta_{\rm extra}{\cal T}_{n}
\ee

Let us compare this to the expected transformation of a bulk scalar field under a special conformal transformation. Again as in the vector case it is useful to understand how the physical operator $\phi_{\rm phys}(z,y_1)$ in the gauge fixed theory is related to the degrees of freedom of the non-gauged fixed theory. For gravity the gauge symmetry is diffeomorphisms, and the question is how to label a particular spacetime point. From the bulk point of view the simplest method is to start at a boundary point (which is invariant under the diffeomorphisms we consider, that fall off quickly enough at infinity), and follow a geodesic orthogonal to the boundary for
a certain proper distance. The distance to the boundary is infinite, but we can regularize this by subtracting an infinite piece which is common to all geodesics or equivalently by putting the boundary at $z=\epsilon$ and later sending $\epsilon \rightarrow 0$.
In this way one can define labels $X,Z$ and a bulk scalar field $\phi(X,Z)$, where $X,Z$ are given by the procedure of starting at some point on the boundary and following an orthogonal geodesic
for a certain proper distance. With this definition the position of the scalar field is invariant under diffeomorphisms.  However $\phi$ is not a local operator since it depends on metric along some path.\footnote{Note that this is not how our smearing function seems to label the bulk operator. The smearing function labels the bulk operator by
specifying the points on the complexified boundary which are spacelike to the bulk point, which is also an invariant notion.}
In holographic gauge these geodesics orthogonal to the boundary have fixed $x_{\mu}$, since the Christoffel symbol $\Gamma_{zz}^{\mu}=0$.  Thus
in holographic gauge we can identify $x = X$, $z = Z$ and $\phi(x,z) = \phi(X,Z)$.

To see how this operator behaves under special conformal transformations, we use the same strategy as the gauge field case. The effect of a special conformal transformation on an operator in holographic gauge can be obtained from a standard special conformal transformation, followed by a coordinate transformation
that restores holographic gauge.
\be
\phi'(z',x')=\phi(z+\epsilon_{z},x_{\mu}+\epsilon_{\mu})
\ee
Here to first order in the parameter of the special conformal transformation $b_{\rho}$ one has  \cite{Kabat:2012hp}
\begin{equation}
\epsilon_{\mu} \sim\frac{1}{N} \int d^d x'' \theta(\sigma z'') \sigma z z'' b^{\alpha}T_{\alpha \mu}, \ \ \ \epsilon_{z}=0
\end{equation}
where the $1/N$ comes from canonically normalizing the kinetic term of the gravity fluctuations. So to first order
\begin{equation}
\phi'(z',x')=\phi(z,x)+\epsilon_{\mu}\partial^{\mu}\phi(z,x).
\end{equation}
This involves a bi-local smearing on the boundary of the operator
$b^{\alpha}T_{\alpha \mu}(x_1)\partial^{\mu}{\cal O}(x_2)$, which can be expanded around 
$b^{\alpha}T_{\alpha \mu}\partial^{\mu}{\cal O}(x)$ using exactly the operators in (\ref{ttrans}). Thus we see that the operators we constructed in the CFT by demanding micro-causality, have the same behavior under special conformal transformations as a bulk scalar field in holographic gauge.

\subsection{Scalar commutator}
We saw that we can add smeared non-primary scalar operators to the definition of a bulk field such that the three-point function $\langle C_{\alpha \mu \beta \nu}(x)\phi(z,y_1){\cal O}(y_2)\rangle$ does not suffer from non-analyticity at bulk space-like separation.  Hence commutators with the bulk field vanish when inserted
inside a three-point function.  That is
\bea
\nonumber
&& \langle C_{\alpha \mu \beta \nu}(x) [\phi(z,y_1),{\cal O}(y_2)]\rangle = 0 \\
&& \langle [C_{\alpha \mu \beta \nu}(x),\phi(z,y_1)]{\cal O}(y_2)\rangle =0
\label{ccom}
\eea
whenever the points in the commutator are spacelike separated. This does not imply that the bulk scalar commutes with the boundary stress
tensor at spacelike separation, for example
$\langle [T_{\mu \nu}(x),\phi(z,y_1)]{\cal O}(y_2)\rangle$ does not need to vanish.  But we claim it does imply that
\begin{equation}
\langle T_{\mu \nu}(x)[\phi(z,y_1),{\cal O}(y_2)]\rangle =0
\label{tcom}
\end{equation}
at bulk spacelike separation to leading order in $1/N$. The argument is similar to the gauge field case. For (\ref{tcom}) to be non-zero the commutator
$[\phi(z,y_1),{\cal O}(y_2)]$ must be a scalar operator that is linear in $T_{\alpha \beta}$.  For example it could be $T_{\mu \nu}(y_1-y_2)^{\mu}(y_1-y_2)^{\nu}$.  But for a conserved stress tensor, given the available operators,
if the commutator is linear in $T_{\alpha\beta}$ then (\ref{ccom}) will not be zero.\footnote{Given a non-conserved spin-two operator one would have operators
available such as its divergence, but in this case one would also be able to construct higher-dimension primary scalars that make the bulk field obey micro-causality.}

While this argument relies on the operator $C_{\alpha \mu \beta \nu}(x)$ with distinct values for all indices, which is only possible in $d\geq 4$, we expect the
results should also hold in $d=3$.  The expressions for the bulk scalar field can simply be analytically continued to $d = 3$.

\section{Higher point functions}
So far we have shown that three-point functions involving a bulk field can be made local to ${\cal O}(1/N)$.  We now use this result to argue that four-point
functions can be made local to ${\cal O}(1/N^2)$.

Consider a four-point function with one bulk scalar operator and three boundary scalar operators. We claim that the four-point function with the bulk operator constructed
as above will obey micro-causality. That is
\be
\langle [\phi(z,x_1), {\cal O}(x_2)] {\cal O}(x_3) {\cal O}(x_4)\rangle =0
\label{4point}
\ee
whenever $(z,x_1)$ and $x_2$ are spacelike separated. To show this we use the OPE of
${\cal O}(x_3)$ and $ {\cal O}(x_4)$
\be
{\cal O}(x_3) {\cal O}(x_4) = \sum c_{i}(x_3,x_4){\cal G}^{i}(x_3)
\ee
where ${\cal G}^{i}(x_3)$ includes a complete set of CFT operators, and we have suppressed any spin indices on ${\cal G}^{i}$. Then
\be
\langle [\phi(z,x_1), {\cal O}(x_2)] {\cal O}(x_3) {\cal O}(x_4)\rangle =\sum c_{i}(x_3,x_4)
\langle [\phi(z,x_1), {\cal O}(x_2)] {\cal G}^{i}(x_3)\rangle 
\label{4point2}
\ee
This reduces the problem to a three-point function.
We have shown that if ${\cal G}^{i}$ is a scalar, then $\phi(z,y_1)$ can be constructed such that the commutator vanishes at spacelike separation
in a three-point function
\cite{Kabat:2011rz,Kabat:2012av}.  We have shown in \cite{Kabat:2012av} and in section 3 that if ${\cal G}^{i}$ is a spin-one current, conserved or not,
then again $\phi(z,y_1)$ can be constructed so that the right hand side of (\ref{4point2}) vanishes at spacelike separation. Finally we showed in section 4 that
the same is true when ${\cal G}$ is the stress tensor. This covers the range of operators we expect to find in the low energy bulk theory,
and strongly suggests that the commutator can be made to vanish at spacelike separation in a three-point function whatever ${\cal G}^{i}$ may be.

This suggests that a bulk scalar field can be constructed in such a way that n-point functions with one bulk operator and $n-1$ boundary operators can be made causal,
at least to leading non-trivial order in the $1/N$ expansion, even in quantum gravity.

\bigskip
\bigskip
\goodbreak
\centerline{\bf Acknowledgements}
\noindent
We are deeply grateful to Stanley Deser for valuable discussions.
DK is supported by U.S.\ National Science Foundation grant PHY-1125915 and by grants from PSC-CUNY.
The work of GL was supported in part by the Israel Science Foundation under Grant No.\ 392/09 and under Grant No. 504/13 and  in part by a grant from the GIF, the German-Israeli Foundation for scientific research and development under grant no. 1156-124.7/2011.

\appendix
\section{Bulk gauge theory\label{appendix:gauge}}
In this appendix we present the canonical formalism for a gauge field coupled to matter, in a way that parallels our
treatment of gravity in section \ref{sect:bulk}.  Although the method is different, the results agree with those obtained
in section 2 of \cite{Kabat:2012av} by following Dirac's procedure.  This both illustrates our method and gives us more confidence in our approach.
It also gives us the opportunity to improve on the boundary conditions which were adopted in \cite{Kabat:2012av}.

For simplicity we consider an abelian gauge field $A_M$ coupled to a complex scalar field $\phi$ with action
\be
S = \int d^{d+1}x \sqrt{-G} \left(-D_M \phi^* D^M \phi - {1 \over 4} F_{MN} F^{MN}\right)
\ee
Here $D_M = \partial_M + i q A_M$, and we work in AdS${}_{d+1}$ with metric
\be
ds^2 = {R^2 \over z^2} \left(-dt^2 + \vert d\vec{x} \vert^2 + dz^2\right)
\ee
The theory can be presented in first-order form.
\bea
\label{FirstOrderGauge}
&& S = \int d^{d+1} x \, \pi_\phi \dot{\phi} + \pi_\phi^* \dot{\phi}^* + \pi_i \dot{A}_i + A_0 \left(\partial_i \pi_i + i q (\pi_\phi \phi - \pi_\phi^* \phi^*)\right) \\
\nonumber
&& \qquad - \left({z \over R}\right)^{d-1} \pi_\phi^* \pi_\phi - {1 \over 2} \left({z \over R}\right)^{d-3} \pi_i \pi_i - \left({R \over z}\right)^{d-1} \vert (\vec{\nabla} + i q \vec{A})\phi\vert^2 - {1 \over 4} \left({R \over z}\right)^{d-3} F_{ij} F_{ij}
\eea
$A_0$ is a Lagrange multiplier enforcing the Gauss constraint
\be
\label{Gauss}
\partial_i \pi_i + i q (\pi_\phi \phi - \pi_\phi^* \phi^*) = 0
\ee
We must also impose a gauge condition.  As in \cite{Kabat:2012av} we adopt the holographic (or axial) gauge condition which sets
\be
\label{AxialGauge}
A_z = 0\,.
\ee
This gives us the right number of degrees of freedom.  The gauge field $A_i$ and its conjugate momentum $\pi_i$ contain $2d$ degrees of freedom.
The constraints (\ref{Gauss}), (\ref{AxialGauge}) kill two phase space degrees of freedom, leaving $2d-2$ phase space degrees of freedom or equivalently
$d-1$ configuration space degrees of freedom.  This is appropriate for a massless spin-1 particle which is a vector under the $SO(d-1)$ little group.

The steps now parallel section \ref{sect:bulk}.  To fix the Lagrange multiplier we require that the gauge fixing condition (\ref{AxialGauge}) is preserved by time evolution.  The equation of motion for $A_z$, obtained by varying the action with respect to $\pi_z$, is
\be
\partial_t A_z = \partial_z A_0 + \left({z \over R}\right)^{d-3} \pi_z
\ee
The left hand side vanishes by the gauge condition, so we get an equation we can solve for $A_0$.\footnote{In the approach of \cite{Kabat:2012av} this
equation was imposed
as an additional gauge condition.}  With suitable boundary conditions as $z \rightarrow 0$ --
to be discussed in more detail below -- the solution is
\be
\label{A_0}
A_0 = \int_0^z dz' - \left({z' \over R}\right)^{d-3} \pi_z
\ee
Next we impose the constraint (\ref{Gauss}) and the gauge condition (\ref{AxialGauge}) on the action (\ref{FirstOrderGauge}).  To take the gauge
condition into account we simply set $A_z = 0$.  Then we solve the Gauss constraint for the conjugate momentum $\pi_z$.  Again imposing suitable
boundary conditions as $z \rightarrow 0$ we have
\be
\label{pi_z}
\pi_z = \int_z^\infty dz' \left(\partial_\hi \pi_\hi + i q (\pi_\phi \phi - \pi_\phi^* \phi^*)\right)
\ee

Let us pause to discuss our boundary conditions in more detail.  With the boundary conditions adopted in (\ref{pi_z}), the electric field $\pi_z$ knows
about all the charge at $z' > z$.  In particular there's no flux through the horizon since $\pi_z \rightarrow 0$ as $z \rightarrow \infty$.  We can write the
solution to the Gauss constraint as (${\bf x} = (\vec{x},z)$)
\bea
\label{pi_z2}
&& \pi_z = \int d^d{\bf x}' \, f({\bf x},{\bf x}') \left(\partial_\hi \pi_\hi + i q (\pi_\phi \phi - \pi_\phi^* \phi^*)\right) \\
&& f({\bf x},{\bf x}') = \delta^{d-1}(x - x') \theta(z'-z)
\eea
Using this in (\ref{A_0}), the solution for $A_0$ becomes
\bea
\label{A_02}
&& A_0 = \int d^d{\bf x}' \, g({\bf x},{\bf x}') \left(\partial_\hi \pi_\hi + i q (\pi_\phi \phi - \pi_\phi^* \phi^*)\right) \\
\nonumber
&& g({\bf x},{\bf x}') = - {1 \over (d-2) R^{d-3}} \delta^{d-1}(x - x') \left(z^{d-2} \theta(z' - z) + (z')^{d-2} \theta(z - z')\right)
\eea
The boundary conditions on $A_0$ have been chosen so that $A_0 \sim z^{d-2}$ as $z \rightarrow 0$, which is the expected behavior for a gauge
field near the boundary of AdS.\footnote{In the AdS/CFT dictionary $A_\mu \sim z^{d-2} j_\mu$ as $z \rightarrow 0$ where $j_\mu$ is identified
with a conserved current in the CFT.  Slightly different boundary conditions were used in section 2 of \cite{Kabat:2012av}.}

Finally we consider the resulting brackets.  Plugging the solution to the constraints back into the action (\ref{FirstOrderGauge}) one obtains an
unconstrained action of the form
\be
\label{GaugeAction}
S = \int d^{d+1} x \, \pi_\phi \dot{\phi} + \pi_\phi^* \dot{\phi}^* + \pi_\hi \dot{A}_\hi - {\cal H}(\phi,\pi_\phi,\phi^*,\pi_\phi^*,A_\hi,\pi_\hi)
\ee
The symplectic form for the physical degrees of freedom $\phi,\pi_\phi,\phi^*,\pi_\phi^*,A_\hi,\pi_\hi$ retains its canonical form, so these
degrees of freedom obey the usual brackets.
\beas
&& \lbrace \pi_{\hat{\imath}}(\x), A_{\hat{\jmath}}(\x') \rbrace = \delta_{\hat{\imath}\hat{\jmath}} \, \delta^d(\x - \x') \qquad \hat{\imath},\hat{\jmath} = 1,\ldots,d-1 \\
&& \lbrace \pi_\phi(\x), \phi(\x') \rbrace = \delta^d(\x-\x') \\
&& \lbrace \pi_\phi^*(\x), \phi^*(\x') \rbrace = \delta^d(\x-\x')
\eeas
However these fields have non-trivial brackets with $A_0$ and $\pi_z$.  These brackets can be calculated from the solutions (\ref{pi_z2}), (\ref{A_02}) by
using the canonical brackets for the physical degrees of freedom.  In this way we obtain
\beas
&& \lbrace A_0(\x), A_{\hat{\imath}}(\x') \rbrace = \partial_{\hat{\imath}} g(\x,\x') \\
&& \lbrace A_0(\x), \phi(\x') \rbrace = i q g(\x,\x') \phi(\x') \\
&& \lbrace A_0(\x), \pi_\phi(\x') \rbrace = - i q g(\x,\x') \pi_\phi(\x') \\[10pt]
&& \lbrace \pi_z(\x), A_{\hat{\imath}}(\x') \rbrace = \partial_{\hat{\imath}} f(\x,\x') \\
&& \lbrace \pi_z(\x), \phi(\x') \rbrace = i q f(\x,\x') \phi(\x') \\
&& \lbrace \pi_z(\x), \pi_\phi(\x') \rbrace = - i q f(\x,\x') \pi_\phi(\x')
\eeas
Aside from the different choice of boundary conditions for the kernel $g$, these results match section 2 of \cite{Kabat:2012av},
where the brackets were obtained following Dirac's procedure for constrained systems.

\section{Symplectic form for matter\label{appendix:symplectic}}
Consider a phase space with coordinates $p_i,q_i,P_I,Q_I$ and suppose there is a first-order Lagrangian of the form
\be
L = p_i \dot{q}_i + f_I(p,q,P,Q) \dot{Q}_I - H(p,q,P,Q)
\ee
This applies to gravity in holographic gauge, with $p,q$ representing matter degrees of freedom and $P,Q$ representing the
physical gravity degrees of freedom $\tilde{\pi}^{ij}$, $\tilde{g}_{ij}$.  Following \cite{Faddeev:1988qp} we introduce the canonical
1-form
\be
a = p_i dq_i + f_I dQ_I
\ee
and the symplectic 2-form $\omega = da$.  The canonical brackets are then given by, for example,
\be
\left\lbrace  p_i,q_j \right\rbrace = - (\omega^{-1})_{ij}
\ee
To compute this we decompose the symplectic form in blocks,
\be
\omega = \left(\begin{array}{cc}
a & b \\ c & d
\end{array}\right)
\ee
where
\bea
&& a = \left(\begin{array}{cc} 0 & \delta_{ij} \\ - \delta_{ij} & 0 \end{array}\right)
\qquad\quad\,\, b = \left(\begin{array}{cc} 0 & {\partial f_J \over \partial p_i} \\ 0 & {\partial f_J \over \partial q_i} \end{array}\right) \\
&& c = \left(\begin{array}{cc} 0 & 0 \\ - {\partial f_I \over \partial p_j} & - {\partial f_I \over \partial q_j} \end{array}\right)
\qquad d = \left(\begin{array}{cc} 0 & {\partial f_J \over \partial P_I} \\ - {\partial f_I \over \partial P_J} & {\partial f_J \over \partial Q_I} - {\partial f_I \over \partial Q_J} \end{array}\right)
\eea
By the blockwise inversion theorem
\be
\left(\begin{array}{cc} A & B \\ C & D \end{array}\right)^{-1} =
\left(\begin{array}{cc} (A - B D^{-1} C)^{-1} & -(A - B D^{-1} C)^{-1} B D^{-1} \\
- D^{-1} C (A - B D^{-1} C)^{-1} & D^{-1} + D^{-1} C (A - B D^{-1} C)^{-1} B D^{-1} \end{array}\right)
\ee
Using this to compute $d^{-1}$ one obtains a matrix of the form
\be
d^{-1} = \left(\begin{array}{cc} \cdot & \cdot \\ \cdot & 0 \end{array}\right)
\ee
Using it a second time to compute $\omega^{-1}$, one finds that the upper left block of $\omega^{-1}$ is
\bea
(a - b d^{-1} c)^{-1} & = & \left(a - \left(\begin{array}{cc} 0 & \cdot \\ 0 & \cdot \end{array}\right)
\left(\begin{array}{cc} \cdot & \cdot \\ \cdot & 0 \end{array}\right)
\left(\begin{array}{cc} 0 & 0 \\ \cdot & \cdot \end{array}\right)\right)^{-1}
\eea
But this reduces to $a^{-1} = \left(\begin{array}{cc} 0 & - \delta_{ij} \\ \delta_{ij} & 0 \end{array}\right)$, which
means the coordinates $p_i,q_i$ obey canonical brackets $\left\lbrace p_i,q_j \right\rbrace = \delta_{ij}$ independent of
the functions $f_I$.

\section{Scalar three-point function\label{appendix:scalar}}

In this section for completeness we present results on the three-point function of one bulk scalar operator and two boundary primary scalar operators.
For additional details see \cite{Kabat:2011rz,Kabat:2012av}. The general form of the correlator, with the appropriate behavior under conformal transformations, is
\begin{eqnarray}
& &\langle \phi_{i}(x,z){\cal O}_{j}(y_1){\cal O}_{k}(y_2)\rangle =c_{ijk}\frac{1}{(y_1 -y_2)^{2\Delta_{j}}}
 \left[\frac{z}{z^2+(x-y_2)^2}\right]^{(\Delta_{k}-\Delta_{j})} \nonumber\\
& &\times \left(\frac{1}{\chi-1}\right)^{\Delta_{0}}F(\Delta_{0},\Delta_{0}-\frac{d}{2}+1,\Delta_{i}-\frac{d}{2}+1,\frac{1}{1-\chi})
\label{3pointbulk}
\end{eqnarray}
where
\begin{equation}
\chi=\frac{[(x-y_1)^2+z^2][(x-y_2)^2+z^2]}{z^2(y_2-y_1)^2}
\end{equation}
and $\Delta_{0}=\frac{1}{2}(\Delta_{i}+\Delta_{j}-\Delta_{k})$.

The analytic structure, which controls the commutator between any two of the operators, is
different in different dimensions. We look for non-analyticity in the region $0 < \chi <1$ where all points are bulk space-like separated from each other, since
any non-analyticity in this region would signal a breakdown in micro-causality. If $d$ is an odd integer we can use the transformation formula for the hypergeometric function
\begin{eqnarray}
F(a,b,c,z)&=&(-z)^{-a}\frac{\Gamma(c)\Gamma(b-a)}{\Gamma(c-a)\Gamma(b)}F(a,a-c+1,a-b+1,\frac{1}{z})\nonumber\\
&+&(-z)^{-b}\frac{\Gamma(c)\Gamma(a-b)}{\Gamma(c-b)\Gamma(a)}F(b,b-c+1,b-a+1,\frac{1}{z})
\end{eqnarray}
The use of this formula in (\ref{3pointbulk}) gives the three-point function an expansion about $\chi = 1$ of the form
\begin{equation}
\frac{1}{(\chi-1)^{\frac{d}{2}}} \sum_{n=0}^\infty a_n (1-\chi)^{n+1}
\end{equation}
Due to the square root branch cut, this implies a non-zero commutator at spacelike separation.

If $d$ is an even integer, the above transformation formula is not valid.  Instead one can use 
\begin{eqnarray}
& &F(a,a+n,c,z)=\frac{\Gamma(c)(-z)^{-a}}{\Gamma(c-a)\Gamma(a+n)}\sum_{k=0}^{n-1}\frac{(n-k-1)!(a)_{k}(1-c+a)_{k}}{k!}(-z)^{-k}\nonumber\\
&+&\frac{\Gamma(c)(-z)^{-a}}{\Gamma(a)\Gamma(c-a-n)}\sum_{k=0}^{\infty}\frac{(a+n)_{k}(1-c+a+n)_{k}}{(n+k)! k!}[\psi(k+1)+\psi(n+k+1)\nonumber \\
&-&\psi(a+n+k)-\psi(c-a-n-k)+\ln (-z)]z^{-n-k}
\end{eqnarray}
where $\psi(x)=\frac{\Gamma^{'}(x)}{\Gamma(x)}$ and $(n)_{k}=\frac{\Gamma(n+k)}{\Gamma(n)}$.
Using this in (\ref{3pointbulk}) we get an expansion about $\chi = 1$ of the form
\begin{equation}
\sum_{k=0}^{d/2} b_{k}(1-\chi)^{-\frac{d}{2}+1+k}+\ln(\chi-1) \sum_{k=0}^{\infty} a_k (1-\chi)^{k}
\end{equation}
Again this results in a non-zero commutator at spacelike separation.

Thus whether $d$ is even or odd one finds a non-zero commutator in the region $0<\chi <1$.  The commutator has an expansion in powers of
$(\chi-1)$. It is important to note that, as can be seen from (\ref{3pointbulk}), for fixed ${\cal O}_{j}$ and ${\cal O}_{k}$ the form of the expansion is independent of
the dimension $\Delta_i$.

We wish to define our bulk operator $\phi_{i}(x,z)$ in order to have the smallest possible commutator with the boundary
operators at spacelike separation.  It should also transform as a bulk scalar under
AdS isometries and have the correct boundary behavior
\begin{equation}
\phi_{i}(x,z) \hspace{1mm} \stackrel{z\rightarrow 0}{\rightarrow} \hspace{1mm} z^{\Delta_{i}} {\cal O}_{i}.
\end{equation}
From the above structure we see that if we have a tower of primary scalar operators ${\cal O}_l$ with dimensions
$\Delta_{l}$, whose three-point functions $\langle {\cal O}_l {\cal O}_{j} {\cal O}_{k}\rangle$ are non-zero, then we can redefine the bulk operator
$\phi_{i}(x,z)$ to have the form
\begin{equation}
\phi_{i}(z,x)=\int dx' K_{\Delta_i}(z,x|x'){\cal O}_i(x')+\sum_{l} a_{l} \int dx' K_{\Delta_{l}}(z,x|x'){\cal O}_l(x')
\end{equation}
Here $K_{\Delta}(z,x|x')$ is the smearing function for a dimension $\Delta$ primary scalar.

Since the form of the singularity is the same for any ${\cal O}_{l}$, we
can choose the coefficients $a_{l}$ to cancel the commutator to any desired order in the expansion
about $\chi = 1$.  If we have an infinite number of suitable
higher-dimension operators we can make the bulk scalar commute at spacelike separation.  Fortunately in the $1/N$ expansion one can construct
the necessary ${\cal O}_{l}$ as multi-trace operators built from products of the ${\cal O}_{k}$ and ${\cal O}_{j}$ together with derivative operators.
If ${\cal O}_{j}$
and ${\cal O}_{k}$ are single-trace operators this procedure begins with a
double-trace operator and thus $a_{l} \sim 1/N$.\footnote{For a general discussion of large-$N$ counting in this context
see p.\ 26 of \cite{Kabat:2011rz}.}
For any three-point function involving $\phi_{i}(z,x)$ we will have to add a different tower of higher dimension operators to our definition of the bulk scalar.
At leading order in the $1/N$ expansion these towers exist and are independent.  It should be possible to correct these operators, order-by-order in the $1/N$
expansion, to preserve micro-causality in the bulk.  But it seems clear that at finite $N$ the required towers of operators cannot exist
and micro-causality will be violated.

\section{Constructing operators\label{appendix:operators}}

In this appendix we give details of the construction of the operators ${\cal V}^{mlk}$ and ${\cal M}^{mlk}$.

\subsection{Spin 1}
The operators ${\cal V}^{mlk}$ are defined by
\begin{equation}
{\cal V}^{mlk}=\frac{1}{N}\sum_{m'+l'+k'=m+l+k}a^{mlk}_{m'l'k'} {\cal A}^{m'l'k'}
\end{equation}
where
\be
{\cal A}^{mlk}=\partial_{\mu_{1}\cdots \mu_{m}}\nabla^{2l}j_{\nu}\partial^{\mu_{1}\cdots \mu_{m}}
\nabla^{2k}\partial^{\nu}{\cal O}_{2}
\ee
and the coefficients $a^{mlk}_{m'l'k'}$ are to be chosen such that
\begin{equation}
\langle F_{\mu \nu}(x) {\cal V}^{mlk}(y_1){\cal O}_2(y_2)\rangle =\frac{\left[(y_1-x)_{\mu}(y_2-x)_{\nu}- \mu \leftrightarrow \nu \right][(x-y_1)_{\alpha}(y_1-y_2)^{\alpha}]^{m}}{(y_1-y_2)^{2\Delta_{2}+2+2m+2k}(x-y_1)^{2\Delta_j+2+2m+2l}}
\label{fvo}
\end{equation}
The operator $j_{\mu}$ has dimension $\Delta_{j}$ and is a conserved current if $\Delta_j=d-1$.
The three-point function is evaluated to leading order in $1/N$ as a factorized product of two-point functions,
thus
\begin{eqnarray}
\langle F_{\mu \nu}(x){\cal A}^{m'l'k'}(y_1){\cal O}_2(y_2)\rangle &=&\langle F_{\mu \nu}(x)\partial_{\mu_{1}\cdots \mu_{m'}}\nabla^{2l'}j_{\nu}(y_1)\rangle \times \nonumber\\
&&\langle \partial^{\mu_{1}\cdots \mu_{m'}}
\nabla^{2k'}\partial^{\nu}{\cal O}_{2}(y_1){\cal O}_2(y_2)\rangle 
\label{fao}
\end{eqnarray}
Let's study the structure of each term. Using
\bea
\langle F_{\mu \nu}(x)j_{\rho}(y_1)\rangle &=&(1-\Delta_j)\frac{\eta_{\nu \rho}(x-y_1)_{\mu}-\eta_{\mu \rho}(x-y_1)_{\nu}}{(x-y_1)^{2\Delta_j +2}} \nonumber\\
\langle {\cal O}(y_1){\cal O}(y_2)\rangle &=&\frac{1}{(y_1-y_2)^{2\Delta}}
\eea
one can easily see that from symmetry consideration and counting derivatives that the right hand side
of (\ref{fao}) is a linear combination of terms appearing on the right hand side of (\ref{fvo}), with
$m+l+k=m'+l'+k'$. So in principle one can just invert this.  We label
\bea
\langle F_{\mu \nu}(x){\cal A}^{mlk}(y_1){\cal O}^{*}(y_2)\rangle  &\equiv& \langle {\cal A}^{mlk}\rangle \nonumber\\
\langle F_{\mu \nu}(x) {\cal V}^{mlk}(y_1){\cal O}^{*}(y_2)\rangle  &\equiv& \langle  {\cal V}^{mlk}\rangle 
\eea
We start with the initial condition (for a conserved current $\Delta_j=d-1$)
\bea
&&\langle {\cal V}^{0lk}\rangle =\frac{1}{\alpha_{lk}}\langle {\cal A}^{0lk}\rangle \nonumber\\
&&\alpha_{lk}=2\Delta(\Delta_{j}-1)\Pi_{i=0}^{l-1}(2\Delta_{j}+2+2i)(2\Delta_{j}-d+2i)\Pi_{r=0}^{k-1}(2\Delta+2-d+2r)(2\Delta+2+2r)\nonumber
\eea
and use the relationships
\bea
&&\langle {\cal V}^{m+1,l,k}\rangle =-\frac{1}{\beta_{1}}\partial^{\rho}_{x}\partial_{\rho, y_2}\langle {\cal V}^{mlk}\rangle +\frac{\beta_{2}}{\beta_{1}}\langle {\cal V}^{m-1,l+1,k+1}\rangle \nonumber\\
&&\langle {\cal A}^{m+1,l,k}\rangle =-\partial^{\rho}_{x}\partial_{\rho, y_2}\langle {\cal A}^{mlk}\rangle \nonumber\\
&&\beta_{1}=(2\Delta+2+2m+2l)(2\Delta+2+2m+2k)\nonumber\\
&&\beta_{2}=m(3+2\Delta_{j}+2\Delta+3m+2l+2k-d)
\eea
to get an iterative procedure to express ${\cal V}^{mlk}$ in terms of ${\cal A}^{m'l'k'}$.

\subsection{Spin 2}
Here we give a similar discussion for spin two.  For simplicity we only consider operators built from a conserved stress tensor.

We would like to find operators ${\cal M}^{mlk}$ that obey
\bea
&&\langle C_{\alpha \beta \gamma \delta}(x) {\cal M}_{mlk}(y_1){\cal O}(y_2)\rangle =\frac{[(y_1-y_2)_{\alpha}(x-y_1)^{\alpha}]^{m}
}{(x-y_1)^{2d+2m+2l+4}(y_1-y_2)^{2\Delta+2m+2k+4}}\nonumber\\
&&\left[(x-y_1)_{\beta}(x-y_1)_{\delta}(x-y_2)_{\alpha}(x-y_2)_{\gamma}-(\gamma \leftrightarrow \delta)-(\beta \leftrightarrow \alpha)+(\gamma \leftrightarrow \delta \ \ \beta \leftrightarrow \alpha)\right]\nonumber
\eea
We do this by writing
\begin{equation}
{\cal M}_{mlk}=\sum_{m+l+k=m'+l'+k'}b^{m'l'k'}_{mlk} {\cal T}_{m'l'k'}
\end{equation}
where
\be
{\cal T}_{m'l'k'}=\partial_{\mu_{1}\cdots \mu_{m'}}\nabla^{2l'}T_{\rho \nu}\partial^{\mu_{1}\cdots \mu_{m'}}
\nabla^{2k'}\partial^{\nu}\partial^{\rho}{\cal O}
\ee
The three-point function is evaluated to leading order in $1/N$ as a factorized product of two-point functions.
\bea
&&\langle C_{\alpha \beta \gamma \delta}(x) {\cal T}_{mlk}(y_1){\cal O}(y_2)\rangle =
\langle C_{\alpha \beta \gamma \delta}(x) \partial_{\mu_{1}\cdots \mu_{m'}}\nabla^{2l'}T_{\rho \nu}(y_1) \rangle \times \nonumber\\
&& \qquad \langle \partial^{\mu_{1}\cdots \mu_{m'}}\nabla^{2k'}\partial^{\nu}\partial^{\rho}{\cal O}(y_1) {\cal O}(y_2) \rangle
\eea
It is useful to note that the only term in the $\langle TT \rangle$ correlator, that contributes to the $\langle CT \rangle$ correlator when all indices on
the Weyl tensor are taken to have distinct values, is
\begin{equation}
\langle T_{\mu \nu}(x_1) T_{\rho \sigma}(y_1)\rangle=d(d-1)(\eta_{\mu \sigma}\eta_{\nu \rho}
+\eta_{\mu \rho}\eta_{\nu \sigma})\frac{1}{(x-y_1)^{2d}}+\cdots.
\end{equation}

We now describe an iterative procedure to get ${\cal M}_{mlk}$.  Let us label
\bea
\langle {\cal M}_{mlk}\rangle  &\equiv&
\langle C_{\alpha \beta \gamma \delta}(x) {\cal M}_{mlk}(y_1){\cal O}(y_2)\rangle \nonumber\\
\langle {\cal T}_{mlk}\rangle  &\equiv&
\langle C_{\alpha \beta \gamma \delta}(x) {\cal T}_{mlk}(y_1){\cal O}(y_2)\rangle
\eea
We start with the initial condition
\bea
&&\langle {\cal M}_{0lk}\rangle =\frac{1}{\beta_{lk}}\langle {\cal T}_{0lk}\rangle  \nonumber\\
&&\beta_{lk}=2d(d-1)\Pi_{i=0}^{k+1}(2d+2i)\Pi_{j=1}^{k}(d+2j)\Pi_{i=0}^{l+1}(2\Delta+2i)\Pi_{j=1}^{l}(2\Delta-d+2j)\nonumber
\eea
and use the relationships
\bea
&&\langle {\cal M}_{m+1,l,k}\rangle =-\frac{1}{\gamma_{2}}\partial^{\rho}_{x}\partial_{\rho, y_2}\langle {\cal M}_{mlk}\rangle +\frac{\gamma_{1}}{\gamma_{2}}\langle {\cal M}_{m-1,l+1,k+1}\rangle \nonumber \\
&&-\partial^{\rho}_{x}\partial_{\rho, y_2}\langle {\cal T}_{mlk}\rangle =\langle {\cal T}_{m+1,l,k}\rangle \nonumber\\
&&\gamma_{1}=m(5+d+3m+2l+2\Delta +2k)\nonumber\\
&&\gamma_{2}=(2d+2m+2l+4)(2\Delta +2m+2k+4) \nonumber
\eea
to compute in an iterative procedure ${\cal M}_{mlk}$ as linear combinations of ${\cal T}_{m'l'k'}$.

\section{Behavior under conformal transformations\label{appendix:conformal}}
In this appendix we give an explicit computation of the behavior under special conformal transformations of the operators ${\cal A}_{n}$ for $n=0$ and $n=1$
in gauge theory, and for ${\cal T}_0$ in gravity.  For gauge theory we use the following transformation rules.  To first order in the parameter $b_{\rho}$
\bea
x'_{\mu}&=&x_{\mu}+2b\cdot x x_{\mu}-b_{\mu}x^2\nonumber\\
\partial^{'}_{\mu}&=&\partial_{\mu}-2b\cdot x \partial_{\mu}-2b_{\mu}x^{\lambda}\partial_{\lambda}+2x_{\mu}b^{\lambda}\partial_{\lambda}\nonumber \\
j'_{\mu}&=&j_{\mu}+2x_{\mu}b \cdot j-2b_{\mu} x \cdot j -2(d-1)b \cdot x j_{\mu} \nonumber\\
{\cal O}'&=&(1-2\Delta b \cdot x){\cal O}\nonumber
\eea
Now ${\cal A}_{0}=j_{\mu}\partial^{\mu}{\cal O}$, and using the above transformation to first order in $b_{\rho}$ one finds \cite{Kabat:2012av}
\bea
( j_{\mu}\partial^{\mu}{\cal O})'=j_{\mu}\partial^{\mu}{\cal O}(1-2b \cdot x (\Delta+d))-2\Delta b \cdot j {\cal O}
\eea
This matches the property advertised in (\ref{vtrans}), that the only terms in $({\cal A}_{0})'$ that differ from a primary scalar involve $b \cdot j$.

Now consider ${\cal A}_{1}$.  From the requirement that it has the correct three-point correlation function with $F_{\mu \nu}$ and ${\cal O}$ we found
in \cite{Kabat:2012av} that
\begin{equation}
{\cal A}_{1}\sim\left(\frac{1}{2d^2}\nabla^{2}j_{\rho}\partial^{\rho}{\cal O}+\frac{1}{2(\Delta+1)(2\Delta+2-d)}j_{\rho}\nabla^{2}\partial^{\rho}{\cal O}-\frac{1}{2d(\Delta+1)}\partial_{\rho}j_{\mu}\partial^{\rho}\partial^{\mu}{\cal O}\right)
\end{equation} 
Now using the above transformation one finds
\bea
(\nabla^{2}j_{\rho}\partial^{\rho}{\cal O})'&=&(\nabla^{2}j_{\rho}\partial^{\rho}{\cal O})(1-2(\Delta+d+2)b\cdot x)+4b^{\rho}\partial_{\mu}j_{\rho}\partial^{\mu}{\cal O}-\nonumber\\
&&2db^{\rho}\partial_{\rho}j_{\mu}\partial^{\mu}{\cal O}-2\Delta b^{\mu}\nabla^{2}j_{\mu}{\cal O}\nonumber\\
(j_{\rho}\nabla^{2}\partial^{\rho}{\cal O})'&=&(j_{\rho}\nabla^{2}\partial^{\rho}{\cal O})(1-2(\Delta+d+2)b\cdot x)-\nonumber\\
&&2(2\Delta+2-d)b^{\rho}j_{\mu}\partial^{\rho}\partial^{\mu}{\cal O}-2(\Delta+2)b^{\mu}j_{|mu}\nabla^2{\cal O}\nonumber\\
(\partial_{\rho}j_{\mu}\partial^{\rho}\partial^{\mu}{\cal O})'&=&(\partial_{\rho}j_{\mu}\partial^{\rho}\partial^{\mu}{\cal O})(1-2(\Delta+d+2)b\cdot x)-2(\Delta+1)b^{\rho}\partial_{\rho}j_{\mu}\partial^{\mu}{\cal O}-\nonumber\\
&&2(\Delta+1)b^{\mu}\partial_{\rho}j_{\mu}\partial^{\rho}{\cal O}+
2b^{\mu}j_{\mu} \nabla^2{\cal O}-2db_{\mu}j_{\rho}\partial^{\mu}\partial^{\rho}{\cal O}
\eea
Putting this all together, again the only terms in $({\cal A}_{1})'$ that differ from a primary scalar are those which involve $b\cdot j$.

For gravity we need the transformation of ${\cal T}_0 = T_{\mu \nu}\partial^{\mu}\partial^{\nu}{\cal O}$.
We have
\begin{eqnarray}
T'_{\mu \nu}&=&T_{\mu \nu}(1-2d(b\cdot x))+2b^{\delta}x_{\nu}T_{\mu \delta}-2b_{\nu}x^{\delta}T_{\mu \delta}+2b^{\delta}x_{\mu}T_{\delta \nu}-2b_{\mu}x^{\delta}T_{\delta \nu}\nonumber\\
(\partial^{\mu}\partial^{\nu}{\cal O})'&=&(\partial^{\mu}\partial^{\nu}{\cal O})(1-2(\Delta+2)b \cdot x)-2(\Delta+1)(b^{\mu}\partial^{\nu}+b^{\mu}\partial^{\nu}){\cal O}-\nonumber\\
&&2b^{\nu}x^{\lambda}\partial_{\lambda}\partial^{\mu}{\cal O}-2b^{\mu}x^{\lambda}\partial_{\lambda}\partial^{\nu}{\cal O}
+2x^{\nu}b^{\lambda}\partial_{\lambda}\partial^{\mu}{\cal O}+2x^{\mu}b^{\lambda}\partial_{\lambda}\partial^{\nu}{\cal O}
\end{eqnarray}
from which we find to first order in $b_{\mu}$
\bea
(T_{\mu \nu}\partial^{\mu}\partial^{\nu}{\cal O})'=(T_{\mu \nu}\partial^{\mu}\partial^{\nu}{\cal O})(1-2(\Delta+2+d)b \cdot x)-4(\Delta+1)b^{\mu}T_{\mu \nu}\partial^{\nu}{\cal O}\nonumber
\eea
Again the terms that differ from a primary scalar have the form (\ref{ttrans}).

\providecommand{\href}[2]{#2}\begingroup\raggedright\endgroup

\end{document}